\pdfoutput=1
\documentclass[atlasdraft=false, cernpreprint,UKenglish, texmf, orcidlogo]{atlasdoc}
\usepackage{atlaspackage}
\usepackage{graphicx}
\usepackage{subfig}
\usepackage{atlasbiblatex}
\usepackage{multirow}
\usepackage{atlasphysics}
\graphicspath{{logos/}{figures/}}

\usepackage{ANA-STDM-2022-15-PAPER-defs}

\AtlasTitle{Measurement and interpretation of inclusive $W\gamma$ production in proton--proton collisions at $\sqrt{s}=13$~TeV using the ATLAS detector}

\AtlasAbstract{%
Differential cross-section measurements are presented for the production of a $W$ boson in association with a photon. The analysis is performed using proton--proton collision data collected by the ATLAS experiment at $\sqrt{s}= 13$~TeV, corresponding to an integrated luminosity of 140~fb$^{-1}$. The differential cross sections are measured in the $W\gamma \rightarrow \ell \nu \gamma$ decay channel ($\ell=e,\mu$) as a function of 16 observables. Collectively, these observables probe the kinematic properties of the $W\gamma$ system, the radiation amplitude zero effect predicted for the $W\gamma$ final state, the polarisation of the $W$ boson, the charge conjugation and parity structure of the $WW\gamma$ triple gauge coupling, and the parton distribution functions of the proton. The data are corrected for the effects of detector inefficiency and resolution and are sufficiently precise that they can be used to distinguish between different state-of-the-art theoretical predictions provided by \sherpa{}, \amc, and \geneva{}. The differential cross sections are used to search for anomalous weak-boson self-interactions induced by dimension-six operators within an effective field theory. For CP-odd operators, dedicated detector-corrected observables based on the outputs of neural networks are found to be particularly sensitive to the interference between the Standard Model and dimension-six scattering amplitudes. Constraints are placed on the Wilson coefficients of the \oW, \oHWB, \oWtil{} and \oHWBtil{} operators in the effective field theory. The sensitivity to the \oHWBtil{} operator is improved by a factor of 2.5 compared to previous measurements in other final states.
}

\author{The ATLAS Collaboration}

\AtlasRefCode{STDM-2022-15}

\PreprintIdNumber{CERN-EP-2026-078}

\AtlasJournal{Eur.~Phys.~J.~C}
\hypersetup{pdftitle={ATLAS document},pdfauthor={The ATLAS Collaboration}}

\begin{document}

\maketitle

\tableofcontents

\section{Introduction}
\label{sec:intro}

\newcommand{\AtlasCoordFootnote}{%
ATLAS uses a right-handed coordinate system with its origin at the nominal interaction point (IP)
in the centre of the detector and the \(z\)-axis along the beam pipe.
The \(x\)-axis points from the IP to the centre of the LHC ring,
and the \(y\)-axis points upwards.
Polar coordinates \((r,\phi)\) are used in the transverse plane,
\(\phi\) being the azimuthal angle around the \(z\)-axis.
The pseudorapidity is defined in terms of the polar angle \(\theta\) as \(\eta = -\ln \tan(\theta/2)\) and is equal to the rapidity
$ y = \frac{1}{2} \ln \left( \frac{E + p_z}{E - p_z} \right) $ in the relativistic limit.
Angular distance is measured in units of \(\Delta R \equiv \sqrt{(\Delta y)^{2} + (\Delta\phi)^{2}}\).}

\begin{figure}[tp]
\centering
\includegraphics[width=0.35\textwidth]{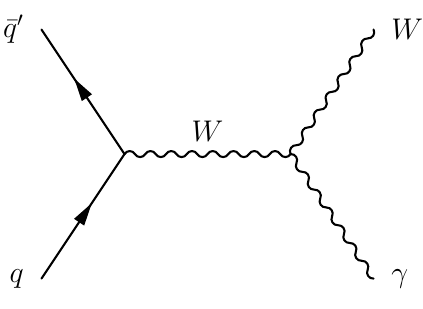} \qquad \qquad
\includegraphics[width=0.35\textwidth]{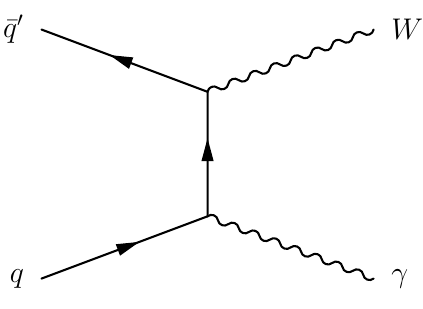}
\caption{$W\gamma$ production in the $s$-channel (left) and $t$-channel (right). The $s$-channel scattering amplitude contains the $WW\gamma$ triple gauge coupling. \label{fig:wy}}
\end{figure}

The production of a $W$ boson in association with a photon in proton--proton collisions is sensitive to a diverse range of physical phenomena. The $s$-channel scattering amplitude contains the $WW\gamma$ triple gauge coupling as shown in Figure~\ref{fig:wy}. Interference between the $s$-channel scattering amplitude and the $t$- and $u$- channel scattering amplitudes results in a complete cancellation at certain points in the phase space \cite{Mikaelian:1979nr,Goebel:1980es,Brodsky:1982sh,Brown:1982xx,Baur:1994sa}, a feature known as radiation amplitude zero. Radiation amplitude zero is an effect that occurs only at leading-order (LO) in the Standard Model (SM) and the effect is removed when either (i) higher-order quantum chromodynamic (QCD) corrections are included, or (ii) when the $WW\gamma$ coupling deviates from the SM prediction \cite{Baur:1994sa}. Measurements of inclusive \wgam{} production can therefore be used to study QCD calculations, to stress-test the gauge structure of the electroweak interaction, and to search for new physics beyond the SM (BSM physics).

Inclusive \wgam{} production is particularly sensitive to the dimension-six operators in effective field theory (EFT) that induce anomalous electroweak-boson self-interactions or anomalous interactions between the electroweak bosons and the Higgs boson. The anomalous interactions induce small deviations from the SM predictions and the largest effects should arise from the interference between the SM scattering amplitude and the BSM scattering amplitude (which contains the anomalous interaction). However, for diboson production, there exists a \textit{no-interference theorem} \cite{Azatov:2016sqh} that states that the interference is suppressed because the dominant contributions to the SM and BSM scattering amplitudes occur for different $W$ boson helicity. Despite this, it has been shown that interference effects can be restored by accessing the off-diagonal elements of the $W$ boson spin density matrix via  either (i) measurements of the $W$ boson decay angles~\cite{Panico:2017frx}, or (ii) measurements of angular correlations between the $W$ boson decay products and the photon~\cite{Degrande:2021zpv,Hall:2022bme}. Measurements of these angular observables in the $W\rightarrow \ell \nu$ decay channel ($\ell=e,\mu$) provide competitive sensitivity to operators that conserve charge conjugation and parity (CP)~\cite{CMS:2021cxr}, and have been proposed as the most sensitive observables for studying CP-violating operators ~\cite{Hall:2022bme}.

Measurements of inclusive \wgam{} production can also be used to test the state-of-the-art predictions provided by analytical QCD calculations and Monte Carlo event generators. Analytical calculations have shown that the next-to-leading-order (NLO) inclusive $W\gamma$ cross section is a factor of three larger than the LO cross section \cite{Baur:1994sa,DeFlorian:2000sg}, with next-to-next-to-leading-order (NNLO) corrections further increasing the cross section by about 20\% \cite{Campbell:2021mlr,Cridge:2021hfr}. Differential cross section measurements of the \wgam{} final state can test these calculations across a wide kinematic region. Alternatively, restricting the quark/gluon radiation in the final state by applying a jet veto suppresses the higher-order contributions and allows the radiation amplitude zero effect to be studied. Finally, inclusive \wgam{} production can also be used to probe the parton distribution functions (PDFs) of the proton. The charge of the lepton from the $W$ boson decay determines whether the dominant contributions to the initial state arise from $u\bar{d}$ or $d\bar{u}$ at leading order. The photon preferentially couples to up-type partons and differential measurements of inclusive \wgam{} boost asymmetries can then help untangle the contributions of the valence quark and sea quark contributions \cite{Yang:2022bxv}.

In this paper, differential cross sections are measured for inclusive \wgam{} production in the $W\rightarrow \ell \nu$ decay channel using proton--proton collision data collected at a centre-of-mass energy $\sqrt{s}=13$ TeV and with an integrated luminosity of $140$~fb$^{-1}$. The measurements are divided into five categories:

\noindent{\textbf{Inclusive \wgam{} kinematic observables}} probe the basic kinematic properties of the \wgam{} system. Differential cross sections are measured as a function of the photon transverse momentum\footnote{\AtlasCoordFootnote}, \pty{}, the charged-lepton transverse momentum, \ptl, the photon pseudo-rapidity, \etay{},
the invariant mass of the lepton-photon system, \mly{}, and the transverse mass of the \wgam{} system, \mtlvy{}~\cite{Baur:1994sa}.

\noindent{\textbf{Radiation amplitude zero}} is probed by measuring the differential cross section as a function of the difference in pseudorapidity between the charged lepton and the photon, \detaly{}.

\begin{figure}[tp]
\centering
\includegraphics[width=0.55\textwidth]{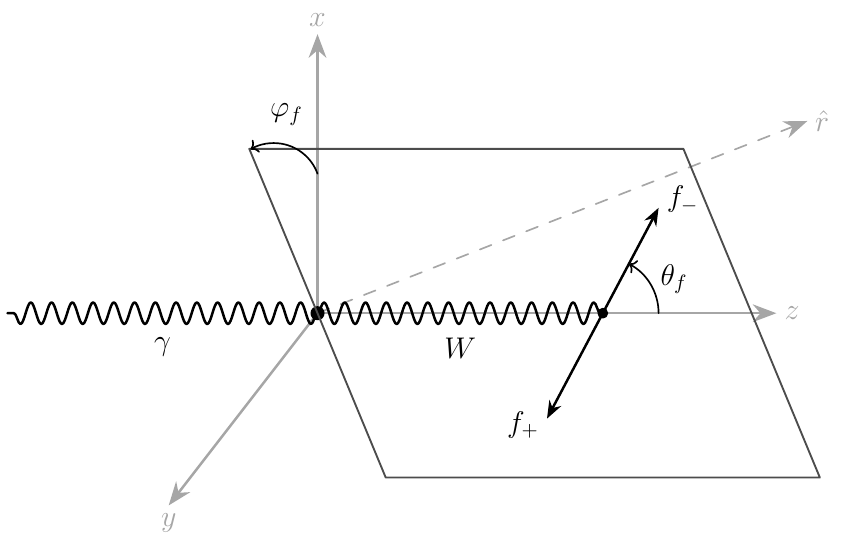}
\caption{The reference frame used to measure the $W\rightarrow \ell\nu$ decay angles, $\theta_f$, and $\phi_f$, is constructed as follows. First, the measured missing transverse momentum is taken as a proxy for the neutrino transverse momentum; the neutrino longitudinal momentum is determined, with a two-fold ambiguity, by assuming a $W\rightarrow \ell\nu$ decay and imposing a $W$ boson mass constraint on the $\ell\nu\gamma$ system. One of the two solutions is chosen at random. Second, the $z$ axis is taken to be the $W$ boson direction in the centre-of-mass frame of the $\ell\nu\gamma$ system. A vector, $\hat{\textbf{ r}}$, is defined in the centre-of-mass frame that defines the boost direction to the lab frame. The $y$ axis is defined by $\hat{\textbf{ y}} = \hat{\textbf{ z}} \times \hat{\textbf{ r}}$ and the $x$ axis is defined as $\hat{\textbf{ x}} = \hat{\textbf{ y}} \times \hat{\textbf{ z}}$. The azimuthal decay angle, $\phi_f$, is taken to be the azimuthal angle of the negative helicity fermion ($f_{-}$) in this special frame. The polar decay angle, $\theta_f$, is defined in the rest frame of the $W$ boson relative to the $W$ boost direction. \label{fig:wy_ref_frame}}
\end{figure}

\noindent{\textbf{Polarisation-sensitive observables}} are constructed from the $W\rightarrow \ell\nu$ decay angles ($\theta_f$, $\phi_f$), which are defined using the negative helicity fermion from the $W$ boson decay and measured in a special reference frame~\cite{Panico:2017frx}, which is shown in Figure~\ref{fig:wy_ref_frame}. A double-differential cross section is measured as a function of $\theta_f$ and $\phi_f$, which directly probes the $W$ boson spin density matrix. In addition, a double differential cross section is measured as a function of $\phi_f$ and $p_{\textrm T}^{\gamma}$, which is sensitive to EFT interference effects that grow with energy.

\noindent{\textbf{CP-sensitive observables}} are designed to be sensitive to the interference effects caused by anomalous CP-violating $WW\gamma$ interactions~\cite{Degrande:2021zpv,Hall:2022bme}. Two types of CP-sensitive differential cross sections are measured. The first is the differential cross section as a function of the pseudorapidity-ordered azimuthal angle between the lepton and the photon, $\Delta\phi_{\ell\gamma} = \phi_i - \phi_j$ (with $\eta_i>\eta_j$). The second is the differential cross section as a function of an optimised observable, \onn{}, which is constructed from the outputs of a multi-class neural network~\cite{Hall:2022bme}. \dphily{} is intrinsically CP-sensitive as a parity-odd observable, whereas \onn{} is trained to optimally separate the constructive- and destructive- interference contributions for a given operator in the effective field theory. The construction of \onn{} is discussed further in Section~\ref{sec:onn}.

\noindent{\textbf{PDF-sensitive observables:}} The boost asymmetry is defined as
\begin{equation*}
A_\textrm{boost}^{W\gamma} = \frac{\sigma_{\gamma}^\textrm{fwd} - \sigma_{\gamma}^\textrm{cent}}{\sigma_{\gamma}^\textrm{fwd} + \sigma_{\gamma}^\textrm{cent}},
\end{equation*}
where $\sigma_{\gamma}^\textrm{fwd}$ is the cross section for events with $|\eta_\gamma|> |\eta_\ell|$ and $\sigma_{\gamma}^\textrm{cent}$ is the cross section for events with $|\eta_\gamma|< |\eta_\ell|$. The boost asymmetry is measured separately for events containing leptons and antileptons and as a function of the photon pseudorapidity and the (anti)lepton pseudorapidity.

Previous measurements of inclusive \wgam{} production have been carried out at the D0 and CDF experiments in proton--antiproton collisions at $\sqrt{s}=1.96$~TeV~\cite{D0:2005djn,D0:2008abl,CDF:2004tuf}. At the LHC, the ATLAS and CMS experiments have measured inclusive $W\gamma$ production in proton--proton collisions at $\sqrt{s}=7$~TeV~\cite{ATLAS:2013way,CMS:2013ryd} . More recently, the CMS experiment has measured the inclusive \wgam{} fiducial and differential cross sections at $\sqrt{s}=13$~TeV, providing the most precise measurements to date and setting tight constraints on anomalous CP-conserving $WW\gamma$ interactions~\cite{CMS:2021foa,CMS:2021cxr}. The measurements presented in this paper go beyond those studies in three ways, by providing (i) new sensitivity to the $W$-boson spin density matrix via the double differential cross sections as a function of $\theta_f$ and $\phi_f$, (ii) new sensitivity to CP-violation via the differential cross sections as a function of $\Delta\phi_{\ell\gamma}$ and \onn{}, (iii) new sensitivity to PDFs via the boost asymmetry. In addition, the differential cross section measurements as a function of the various kinematic observables has comparable precision to that achieved previously.


%
\section{ATLAS detector}
\label{sec:detector}


The ATLAS detector~\cite{PERF-2007-01} at the LHC covers nearly the entire solid angle around the collision point.
It consists of an inner tracking detector surrounded by a thin superconducting solenoid, electromagnetic and hadronic calorimeters,
and a muon spectrometer incorporating three large superconducting air-core toroidal magnets.

The inner-detector system (ID) is immersed in a \qty{2}{\tesla} axial magnetic field
and provides charged-particle tracking in the range \(|\eta| < 2.5\).
The high-granularity silicon pixel detector covers the vertex region and typically provides four measurements per track,
the first hit generally being in the insertable B-layer (IBL) installed before Run~2~\cite{ATLAS-TDR-19,PIX-2018-001}.
It is followed by the SemiConductor Tracker (SCT), which usually provides eight measurements per track.
These silicon detectors are complemented by the transition radiation tracker (TRT),
which enables radially extended track reconstruction up to \(|\eta| = 2.0\).
The TRT also provides electron identification information
based on the fraction of hits (typically 30 in total) above a higher energy-deposit threshold corresponding to transition radiation.

The calorimeter system covers the pseudorapidity range \(|\eta| < 4.9\).
Within the region \(|\eta|< 3.2\), electromagnetic calorimetry is provided by barrel and
endcap high-granularity lead/liquid-argon (LAr) calorimeters,
with an additional thin LAr presampler covering \(|\eta| < 1.8\)
to correct for energy loss in material upstream of the calorimeters.
Hadronic calorimetry is provided by the steel/scintillator-tile calorimeter,
segmented into three barrel structures within \(|\eta| < 1.7\), and two copper/LAr hadronic endcap calorimeters.
The solid angle coverage is completed with forward copper/LAr and tungsten/LAr calorimeter modules
optimised for electromagnetic and hadronic energy measurements, respectively.

The muon spectrometer (MS) comprises separate trigger and
high-precision tracking chambers measuring the deflection of muons in a magnetic field generated by the superconducting air-core toroidal magnets.
The field integral of the toroids ranges between \num{2.0} and \qty{6.0}{\tesla\metre}
across most of the detector.
Three layers of precision chambers, each consisting of layers of monitored drift tubes, cover the region \(|\eta| < 2.7\),
complemented by cathode-strip chambers in the forward region, where the background is highest.
The muon trigger system covers the range \(|\eta| < 2.4\) with resistive-plate chambers in the barrel, and thin-gap chambers in the endcap regions.

The luminosity is measured mainly by the LUCID--2~\cite{LUCID2} detector that records Cherenkov light produced in the quartz windows of photomultipliers located close to the beampipe.

Events were selected by the first-level trigger system implemented in custom hardware,
followed by selections made by algorithms implemented in software in the high-level trigger~\cite{TRIG-2016-01,ATLAS:2020esi}.
The first-level trigger accepted events from the \qty{40}{\MHz} bunch crossings at a rate close to \qty{100}{\kHz},
which the high-level trigger further reduced in order to record complete events to disk at about \qty{1.25}{\kHz}.

A software suite~\cite{ATL-SOFT-PUB-2021-001} is used in data simulation, in the reconstruction
and analysis of real and simulated data, in detector operations, and in the trigger and data acquisition
systems of the experiment.


%
\section{Monte Carlo simulation}
\label{sec:datamc}

The inclusive $W\gamma \rightarrow \ell\nu\gamma$ ($\ell=e,\mu$) signal is modelled using two different Monte Carlo (MC) event generators, \sherpa{} and \amc{}. The first sample is generated using \sherpa~2.2.11~\cite{Bothmann:2019yzt} with matrix elements at NLO accuracy in perturbative QCD for up to one additional parton in the final state and at LO accuracy for up to four additional partons. The matrix element calculations were matched
and merged with the \sherpa{} parton shower based on Catani--Seymour dipole factorisation~\cite{Gleisberg:2008fv,Schumann:2007mg} using the MEPS@NLO
prescription~\cite{Hoeche:2011fd,Hoeche:2012yf,Catani:2001cc,Hoeche:2009rj}. The virtual QCD corrections were provided by the \textsc{OpenLoops} library~\cite{Buccioni:2019sur,Cascioli:2011va,Denner:2016kdg}. The \NNPDF[3.0nnlo]~\cite{Ball:2014uwa} set of parton distribution functions were used, along with the dedicated set of tuned parton-shower parameters developed by the \sherpa{} authors. The second inclusive \wgam{} sample is generated using \amc{}~\cite{Alwall:2014hca, Frederix:2018nkq, Frederix:2018nkq-err, Frederix:2012ps}. This sample uses matrix elements at NLO accuracy in perturbative QCD for up to one additional parton in the final state. The PDF set used in the calculation is \NNPDF[3.0nlo]. The matrix elements are matched to the \pythia{} parton shower \cite{Sjostrand:2014zea} using the  \FXFX scheme~\cite{Frederix:2012ps}. \pythia{} is also used to model the hadronisation and underlying event, with the A14 set of tuned parameters~\cite{ATL-PHYS-PUB-2014-021}. This sample is referred to as \mgpy{}. In both the \sherpa{} and \mgpy{} samples, the Frixione isolation parameters for fragmentation photons~\cite{Frixione:1998jh} are set to $\delta_0 = 0.1$, $\epsilon =0.1$ and $n=2$.

The electroweak production of dijets in association with the \wgam{} system constitutes approximately 1\% of the signal process and is not included in the samples discussed above. A sample of these EW $W\gamma \! j \! j$ events is produced with \sherpa{} 2.2.12 and added to the signal predictions. The sample is generated using matrix elements with up to one additional parton at LO in perturbative QCD. The PDF set used in the calculation is \NNPDF[3.0nnlo]. The matrix elements are matched to the \SHERPA parton shower using the MEPS@NLO prescription. The Frixione isolation parameters are $\delta_0 = 0.1$, $\epsilon =0.1$ and $n=2$.

The production of $W\gamma$ with the $W$ boson decaying first to a $\tau$-lepton and then subsequently to an electron or muon is treated as a background in this measurement. Samples of these events are produced using \sherpa{} and \mgpy{} in the same way as for the signal process (including the EW $W\gamma \! j \! j$ contribution). $Z\gamma$ events also act as a source of background when one of the leptons from the $Z$ boson decay is not reconstructed by the detector. An inclusive $Z\gamma$ sample is generated using \sherpa{} 2.2.11 using the same settings as for inclusive \wgam{} production. The $Z\gamma$ sample is scaled by a factor of 1.08 to match the signal strength reported in recent ATLAS measurements~\cite{ATLAS:2022wnf}. A sample of EW $Z\gamma \! j \! j$ events is produced with \sherpa{} 2.2.12 and added to the prompt background, using the same event generator settings as for EW $W\gamma \! j \! j$ events.

The production of $t\bar{t}\gamma$, $tW\gamma$ and $tq\gamma$ is another source of prompt background. These events are modelled using the \mg{} event generator at LO accuracy in QCD, with the \NNPDF[2.3lo]~\cite{Ball:2012cx} PDF sets. The events were interfaced with \pythia{} to model the parton shower, hadronisation, and underlying event, using the A14 set of tuned parameters~\cite{ATL-PHYS-PUB-2014-021}. For all backgrounds with prompt photons, the Frixione isolation parameters are defined in the same way as for the signal samples. The  $t\bar{t}\gamma$ sample is scaled by a factor of 1.44 to match the signal strength reported in recent ATLAS measurements~\cite{ ATLAS:2024hmk,ATLAS:2022wnf}.

Diphoton production acts as background when one of the photons is misidentified as an electron. Samples of these events are produced using \sherpa{} 2.2.4 with matrix elements accurate to NLO in QCD for zero or one partons in the final state and LO in QCD for up to three additional partons. The \NNPDF[3.0nnlo] PDF set was used and the matrix elements were matched to the \sherpa{} parton shower using the MEPS@NLO prescription.

Background processes that do not contain prompt leptons or prompt photons are estimated using data-driven techniques as outlined in Section~\ref{sec:backgrounds}. The data-driven method to determine  the background from $Z\rightarrow e^+e^-$ events in which one electron is reconstructed as a photon (referred to as $e\rightarrow\gamma$) applies scale factors to correct MC simulations.
A $Z+\mathrm{jets}$ sample is produced using \sherpa{} 2.2.11 at NLO accuracy in QCD for up to two additional partons in the final state and at LO accuracy for up to 5 additional partons in the final state. \NNPDF[3.0nnlo] set of parton distributions are used and the MEPS@NLO prescription is used to match the matrix elements to the parton shower. The data-driven estimate to determine the background from events in which a jet is misidentified as a lepton ($j\rightarrow \ell$) makes use of a dijet control region to estimate the $j\rightarrow \ell$ efficiency. Backgrounds in this control region arise from $W+\mathrm{jets}$, $Z+\mathrm{jets}$ and $\gamma+\mathrm{jets}$ production and are subtracted from the data. The  $W+\mathrm{jets}$ sample is produced using the same settings as for $Z+\mathrm{jets}$ production. The $\gamma+\mathrm{jets}$ sample is produced using \sherpa{} 2.2.2 with matrix elements accurate to NLO in QCD for up to one parton in the final state and to LO in QCD for up to four additional partons. \NNPDF[3.0nnlo] set of parton distributions are used and the MEPS@NLO prescription is used to match the matrix elements to the parton shower.

EFT studies are facilitated using dedicated samples produced using the \mg{} event generator at LO accuracy in QCD for up to one additional parton in the final state. The  \NNPDF[2.3lo] PDFs are used in the calculation. \pythia{} is used for parton showering, hadronisation and underlying event with the A14 set of tuned parameters. The matrix elements are matched to the \pythia{} parton shower using the \textsc{CKKW-L} procedure~\cite{Lonnblad:2001iq,Lonnblad:2011xx}. The anomalous coupling effects predicted by EFT operators in the Warsaw basis are implemented in \textsc{SMEFTSim} 3.0~\cite{Brivio:2017btx,Brivio:2020onw}. Samples are produced for the \oWtil{}, \oHWBtil{}, \oW{} and \oHWB{} operators. The $M_W$ scheme is chosen for the EW sector in the \textsc{SMEFTSim} configuration and a $U(2)^3$ symmetry is adopted for light quark transformations. The masses and Yukawa couplings of all fermions except top and bottom quark are set to zero and the CKM matrix is set to the identity matrix.

All MC samples are passed through a detailed simulation of the ATLAS detector~\cite{SOFT-2010-01} based on \textsc{Geant4}~\cite{Agostinelli:2002hh}. The effect of multiple proton--proton interactions (pile-up) in each bunch crossing is accounted for using simulations of inelastic $pp$ collisions produced with \pythia{}~\cite{ATL-PHYS-PUB-2016-017}. These inelastic $pp$ interactions are added to the signal and background simulations and weighted to ensure that the distribution of the average number of pile-up interactions matches that observed in the data. Residual differences between lepton trigger, identification and isolation efficiencies are corrected on an event-by-event basis using \pt{} and $\eta$ dependent scale factors. A similar approach is adopted to account for residual differences in the photon identification and isolation efficiencies.


%
\section{Event and object selection}
\label{sec:evsel}

The measurements are performed using proton--proton collision data collected at a centre-of-mass energy of $\sqrt{s}=13$~TeV, with an integrated luminosity of 140~\ifb~\cite{DAPR-2021-01}. The events are required to pass unprescaled single-lepton triggers~\cite{TRIG-2018-01,TRIG-2018-05}. The minimum transverse momentum thresholds applied to leptons in the trigger depend on the lepton flavour, lepton isolation criteria, and data-taking period, and vary between $20-26$~GeV. Events are  required to satisfy stringent data quality criteria, which ensure that the ATLAS detector was fully operational during stable beam conditions~\cite{DAPR-2018-01}. Events are also required to have at least one reconstructed interaction vertex~\cite{ATLAS:2015jge}, where each candidate vertex has at least two charged-particle tracks with $p_\textrm{T} > 500$~MeV. The primary vertex is defined as the vertex with the highest scalar sum of track $p_\textrm{T}^2$. %

Muons are reconstructed by matching tracks in the inner detector to tracks in the muon spectrometer and have $p_\textrm{T} > 7$~GeV and $|\eta|<2.5$. Muons are required to pass the Loose identification working point and the PflowLoose\_VarRad isolation working point~\cite{MUON-2018-03}. In addition, muons must originate from the primary vertex by requiring $|d_0/\sigma_{d_0}|<3$ and $|z_0 \, \textrm{sin}\theta|<0.5$~mm, where $d_0$ is the
distance of closest approach to the primary vertex in the transverse plane, $\sigma_{d_0}$ is the corresponding uncertainty, and $z_0$
is the longitudinal distance between the primary vertex and the point at which $d_0$ is defined.

Electrons are reconstructed from clusters of energy deposits in the electromagnetic calorimeter matched to tracks reconstructed in the inner detector and have $p_\textrm{T} > 7$~GeV and $|\eta|<2.47$ (but excluding a calorimeter crack region $1.37<|\eta|<1.52$). Electrons are required to satisfy the Loose identification working point~\cite{EGAM-2018-01}, the Loose\_VarRad isolation working point~\cite{EGAM-2021-01}, $|d_0/\sigma_{d_0}|<5$, and $|z_0 \, \textrm{sin}\theta|<0.5$~mm.

Photons are also reconstructed from clusters of energy deposits in the electromagnetic calorimeter and are subsequently defined as converted or unconverted. Converted photons are clusters that are matched to one or two inner detector tracks commensurate with a $\gamma\rightarrow e^+e^-$ conversion vertex, whereas unconverted photons are required to be unmatched to both inner detector tracks and photon conversion vertices. Photons are required to have $p_\textrm{T} > 12$~GeV and $|\eta|<2.37$, excluding the crack region $1.37<|\eta|<1.52$. In addition, photons are required to satisfy the Loose identification working point and the Loose isolation working point~\cite{EGAM-2018-01}.

Jets are reconstructed using the anti-$k_t$ algorithm~\cite{Cacciari:2008gp,Cacciari:2011ma} with a distance parameter of $R=0.4$. The inputs to the jet algorithm are calorimeter topoclusters and inner detector tracks, which are combined using particle-flow algorithms~\cite{PERF-2015-09}. The jet energy scale and resolutions are calibrated using $p_\textrm{T}$- and $|\eta|$- dependent correction factors determined from MC simulations and data-driven methods~\cite{ATLAS:2020cli}. Jets are required to have $p_\textrm{T}>20$~GeV and $|\eta|<4.4$. Jets are also required to pass the tight working point of the jet vertex tagging algorithm~\cite{PERF-2014-03}, which ensures that the jets originate from the primary vertex. Jets containing a $b$-hadron are identified with the DL1r tagger~\cite{FTAG-2019-07} with an average efficiency of 70\% for true $b$-jets, as estimated in ttbar events.

The magnitude of the missing transverse momentum, \ptmiss{}, and the associated azimuthal direction, are calculated from the negative vector sum of the transverse momenta of the muons, electrons, photons, jets, and a track-based soft term~\cite{JETM-2020-03}. The soft term is calculated from all inner-detector tracks that are not matched to any of the reconstructed muons, electrons, photons or jets. %

An object overlap removal procedure is applied to remove ambiguities in the object reconstruction.
Electrons are removed if they share an ID track with a muon. Jets are removed if they have $\Delta R<0.2$ with respect to an electron or $\Delta R<0.4$ with respect to a photon. Jets are also removed if they have less than three tracks and have $\Delta R<0.2$ with respect to a muon. Photons are rejected if they have $\Delta R<0.4$ with respect to an electron or a muon. Electrons and muons are then removed if they have $\Delta R<0.4$ with respect to a jet.

Following the overlap removal, events are required to contain exactly one lepton, at least one photon, and have $p_\textrm{T}^\textrm{miss}>40$~GeV. The leading photon and the lepton are required to have $\pty>30$~GeV and $\ptl>30$~GeV, respectively. The photon is then required to pass the Tight identification working point and the TightCaloOnly isolation working point~\cite{EGAM-2018-01}. The identification and isolation criteria for the lepton are also tightened; the Tight identification working point and Tight\_VarRad isolation working point are required for electrons, whereas the Tight identification working point and the PflowTight\_VarRad  isolation working point are required for muons. Events are rejected if they contain any $b$-jets with $p_\textrm{T}>20$~GeV and $|\eta|<2.5$, to reject backgrounds containing one or
more top quarks. Similarly, events are rejected if the invariant mass of the lepton-photon system is in the range $81<m_{\ell\gamma}<101$~GeV, to reject backgrounds arising from $Z\rightarrow e^+ e^-$ with one of the electrons reconstructed as a photon. Finally, the separation of the lepton and the photon is required to satisfy $\Delta R> 0.8$. The majority of the differential distributions are measured in this \textit{baseline} region. The exceptions are the differential cross sections as a function of \dphily{} and \detaly{}, which are measured in a \textit{jet veto} region in which events are required to have no jets with $p_\textrm{T}>30$~GeV, in addition to the selections that define the \textit{baseline} region.  The double differential cross section as a function of $\phi_f$ and $p_{\textrm T}^{\gamma}$ is measured in both the \textit{baseline} and \textit{jet veto} regions.

Figure~\ref{fig:dataMCcomparison} shows the number of events measured as a function of \pty{}, \etay{}, \mly{} and \detaly{}. The data are compared to a combined prediction containing the signal and all background processes. The signal and prompt backgrounds are modelled using the MC simulations discussed in Sec.~\ref{sec:datamc}. The other backgrounds are determined using the data-driven methods discussed in Sec.~\ref{sec:backgrounds}.
The data are in reasonable agreement with the combined predictions for the \pty{}, \etay{} and \mly{} distributions, which are measured in the \textit{baseline} region. The combined prediction is below the data for the \detaly{} distribution, which is measured in the \textit{jet veto} region. The theoretical uncertainties on the signal and prompt backgrounds are much smaller in the \textit{jet veto} region and it is possible that they are underestimated due to the application of a jet veto~\cite{Stewart:2011cf}.

The number of events selected in the data is 264094, with background processes accounting for 38\% of the yield. The dominant source of background is $j \to \gamma$ background (40\% of total background), followed by prompt backgrounds (27\%), the $\ell \to \gamma$ background (18\%), the $ j \to \ell$ background (8\%), and the pile-up $\gamma$  background (7\%). The agreement between data and combined prediction has also been checked separately in the electron and muon channels, with good agreement and similar signal strengths observed in each channel.

\begin{figure}[tp]
\subfloat[]{\includegraphics[width=0.5\textwidth]{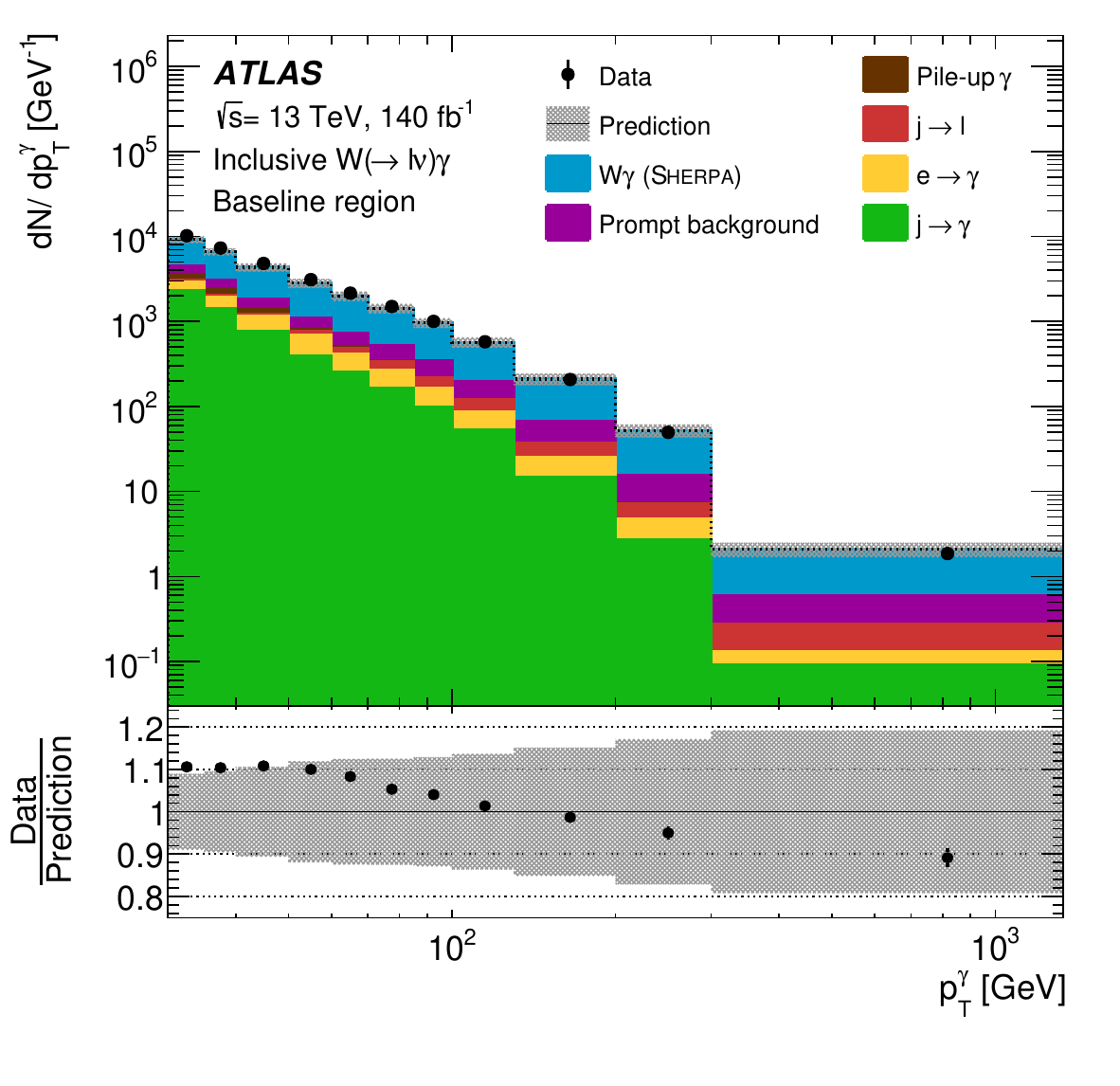}}
\subfloat[]{\includegraphics[width=0.5\textwidth]{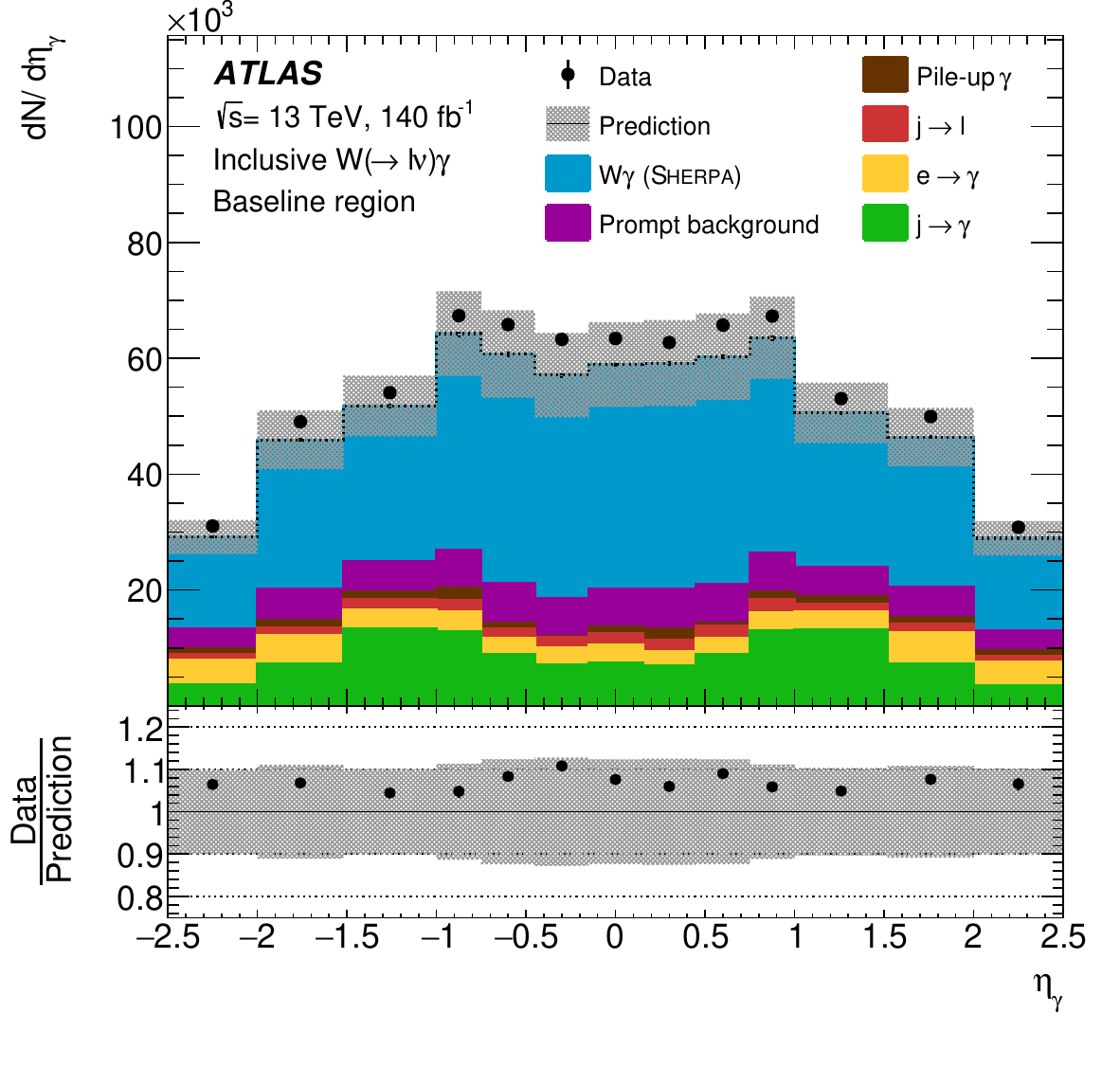}}\\
\subfloat[]{\includegraphics[width=0.5\textwidth]{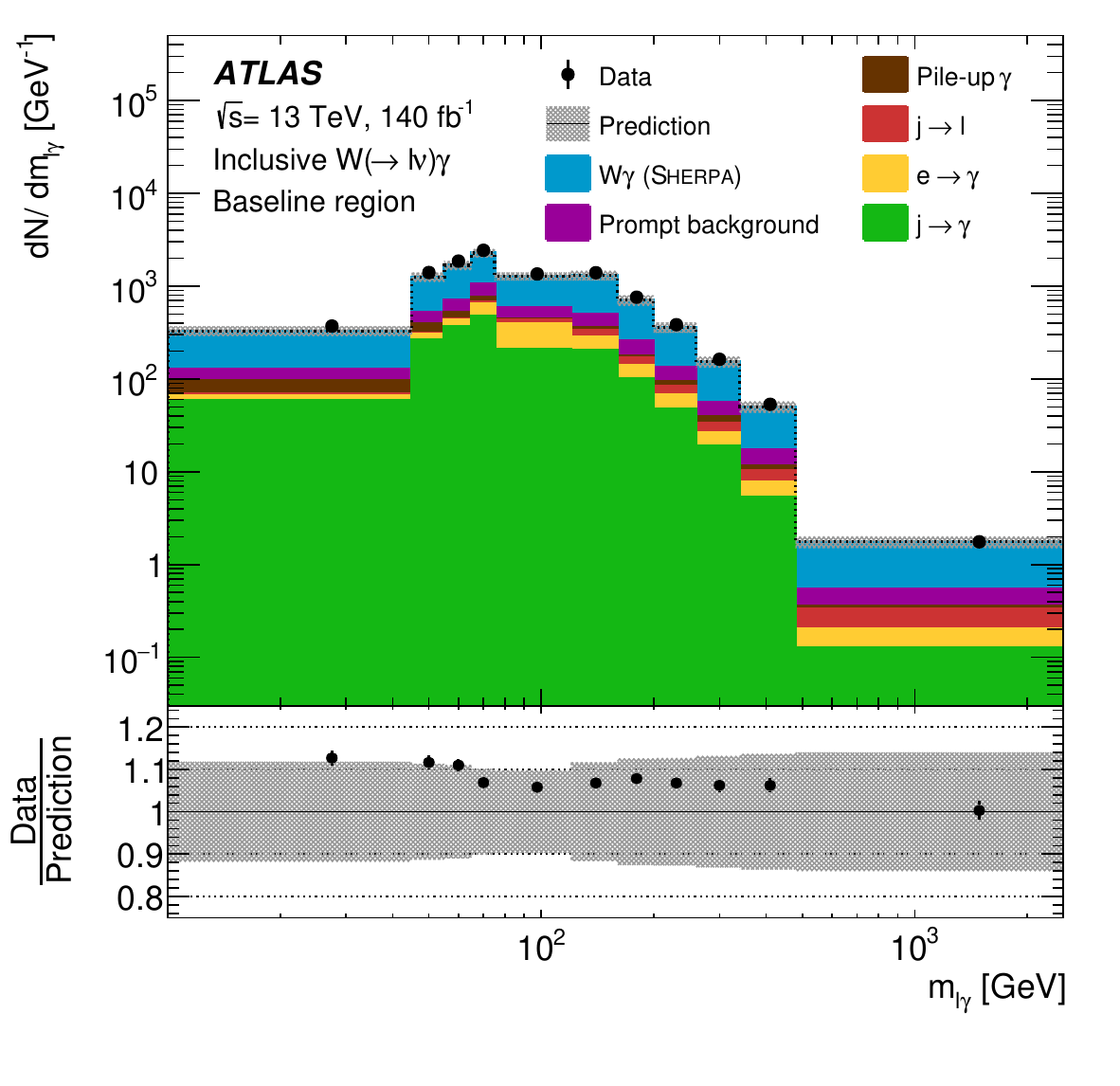}}
\subfloat[]{\includegraphics[width=0.5\textwidth]{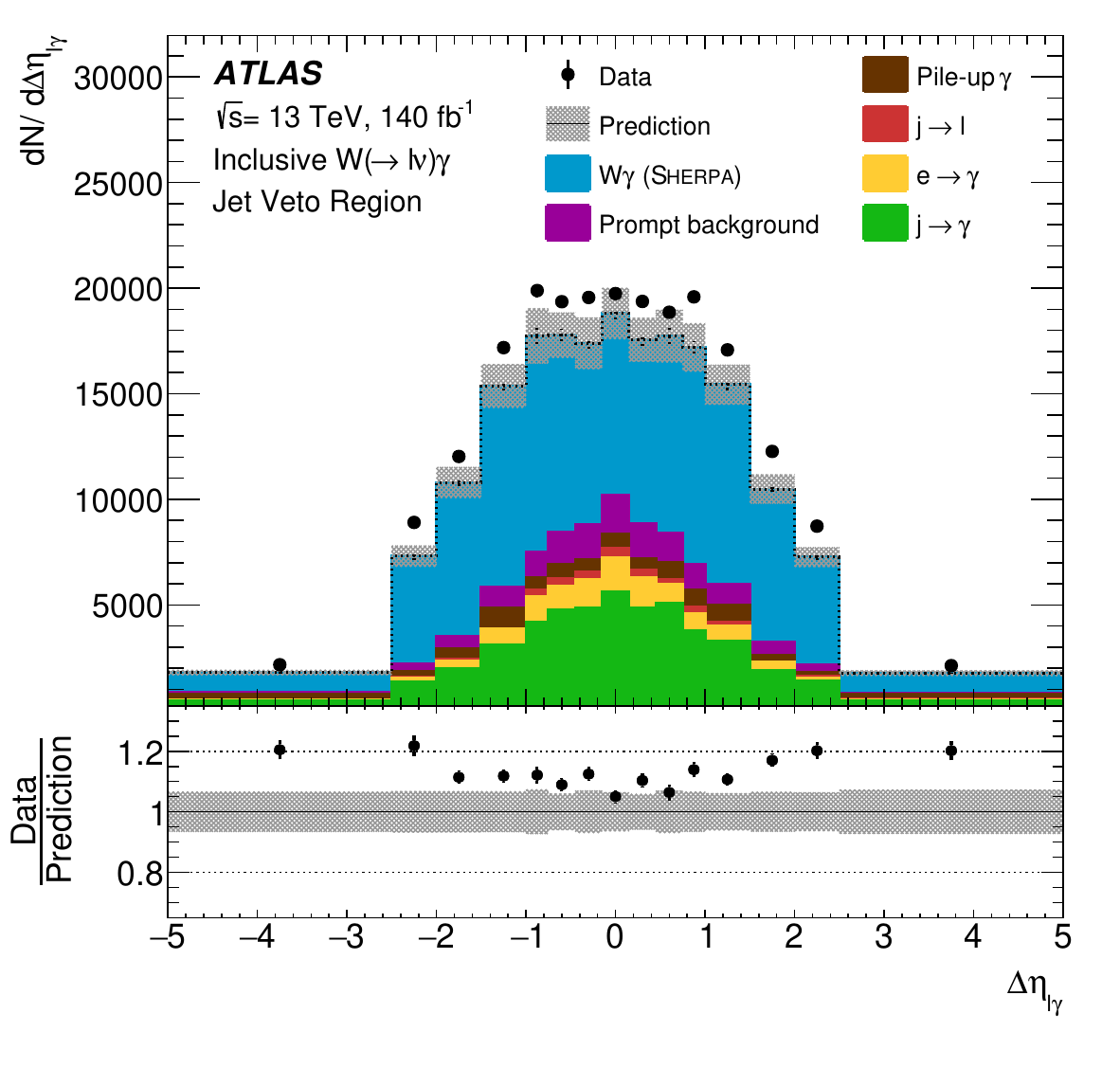}}%
\caption{Predicted and observed yields as a function of (a) \pty{}, (b) \etay{}, (c) \mly{} and (d) \detaly{}. The data are represented as black points and the associated error bars represent the statistical uncertainty.
Background processes with prompt leptons and photons, such as $Z\gamma$ production and the fully leptonic decays of $t\bar{t}\gamma$ production, are estimated by using simulations and labelled as `Prompt.'   Backgrounds arising from non-prompt leptons ($j\rightarrow \ell$), non-prompt photons ($j\rightarrow \gamma$), electrons-faking-photons ($e\rightarrow \gamma$) and pile-up photons are estimated using data-driven techniques as outlined in Section~\ref{sec:backgrounds}.   The total uncertainty on the combined signal and background prediction is shown as a grey band (the calculation of these uncertainties is outlined in Section~\ref{sec:systematics}).
\label{fig:dataMCcomparison}}
\end{figure}


%
\section{Background estimation}
\label{sec:backgrounds}

The backgrounds relevant to this analysis can be grouped into three categories.
The first category comprises prompt backgrounds that arise from processes that produce genuine photons and leptons
in the final state, such as $W(\to\tau\nu)\gamma$, $Z\gamma$, $t\bar{t}\gamma$, $tW\gamma$, and $tq\gamma$.
These contributions are estimated from Monte Carlo simulation as discussed in Section~\ref{sec:datamc}. The theoretical and experimental
uncertainties on these backgrounds are discussed in Section \ref{sec:systematics}.

The second category consists of non-prompt or fake backgrounds, where a reconstructed object is misidentified as a photon or a lepton.
These contributions are not reliably modelled in simulation and are therefore determined with data-driven techniques
based on template fits, control regions, and sideband methods, with associated systematic
uncertainties derived from these procedures.

Finally, the pile-up background arises when a genuine photon from an additional proton--proton
interaction in the same bunch crossing is combined with a lepton from the hard-scatter $W$ boson decay.
Although the photon is real, the combination fakes the $W\gamma$ signal. This background is estimated
separately using data-driven techniques that exploit the separation between the photon and
the primary vertex along the beam axis.

The remainder of this section describes the estimation of the non-prompt and pile-up
backgrounds, which together constitute the dominant reducible contributions to the background of
the analysis.

\subsection{Jets faking photons}

The dominant background for this analysis is due to $W$+jets events where either a jet is misidentified
as a photon or a real photon produced within a hadronic jet passes the photon selection criteria. These
contributions, referred to  as $j \to \gamma$, are estimated using a data-driven method.
It is based on template fits to the photon isolation distribution, and validated using an
independent sideband (ABCD) method~\cite{ATLAS:2010uco}. The measurement of the boost asymmetry uses an independent estimation of this background that follows the same template-fit strategy.

Photon candidates in the analysis are required to pass tight identification and isolation criteria.
The calorimeter isolation variable, defined from the transverse energy within a cone of size $\Delta R=0.4$
around the photon axis, is effective at separating real photons from jets. Prompt photons
produce a narrow peak near zero isolation, while jets give rise to a broad tail at higher isolation values.
This separation is exploited by constructing isolation templates for signal and background.

The signal template is obtained from simulated $W\gamma$ events containing a Tight photon. The background template is derived from data using events in which the photon passes a Loose identification working point but fails the Tight requirement, thereby enriching the sample in jet-induced photons. A small signal leakage into this control sample is estimated using simulation and corrected for in the template construction.
A binned fit to the isolation distribution then determines the relative contributions of real photons and jet-induced photons. The $j\rightarrow \gamma$ background yield is obtained by integrating the fitted jet-induced photon template within the isolation range used in
the analysis. This procedure is performed bin-by-bin for each observable relevant to the differential cross section measurements.

An independent cross-check is performed using an ABCD method, in which photon
identification and isolation are treated as approximately uncorrelated variables. Four regions are defined:
the signal region (tight ID, isolated), and three control regions corresponding to either looser
identification or inverted isolation. The number of jet-induced photons in the signal region is then
predicted from the control regions after correcting for signal leakage and prompt-photon contamination.
The method provides good statistical precision and a conceptually simple validation of the template fit.
Some discrepancies are observed in the highest \pt{} and most forward bins in $\eta$, where correlations between
the two variables are significant, and the ABCD method is expected to become less reliable.

The dominant sources of systematic uncertainty in the $j\rightarrow \gamma$ estimate are associated with the construction of the background template and with the modelling of the prompt photon template.
For the systematic uncertainty associated with the background template, alternative loose identification working points are used to construct the background templates. The spread of the
$j\rightarrow \gamma$ predictions obtained from fitting the isolation distribution with different background templates is taken as the systematic uncertainty, which is typically around 12\% . For the systematic associated with the modelling of the prompt photon template, different event generators
(\sherpa{} and \mgpy{}) are used in the fit to the isolation distribution and the observed differences of around 4\% are propagated as uncertainties.

\subsection{Jets faking leptons}

Events in which a jet is misidentified as a lepton, or where a non-prompt
lepton is produced inside a hadronic shower and satisfies the analysis selection, constitute another source of
background for this analysis.  In the electron channel, such contributions originate mainly from Dalitz decays
or misidentified hadrons, while in the muon channel, they arise predominantly from semileptonic decays of heavy-flavour hadrons. Since the
modelling of these processes is affected by large uncertainties in hadronisation and the detector response, a
fully data-driven strategy is adopted for each lepton channel.

Two complementary approaches are considered: the matrix method, extended with a Poisson likelihood fit,
and the fake factor method~\cite{ATLAS:2022swp}. Both rely on defining two nested lepton selections. ``Baseline'' leptons
satisfy looser identification and isolation criteria, chosen to enhance the contribution of fakes, while
``tight'' leptons correspond to the signal definition used in the analysis. By construction, every tight
lepton is also a baseline lepton. The probability for a baseline lepton to pass the tight requirements is
measured separately in samples enriched in prompt leptons, providing the \emph{real efficiency}
$\epsilon_{r}$, and in samples enriched in fake leptons, giving the \emph{fake efficiency} $\epsilon_{f}$.

The real efficiency is measured using $Z\to\ell\ell$ events. Events with exactly two oppositely charged,
same-flavour leptons with invariant mass within 10~GeV of the $Z$ boson mass are selected. To reduce
diboson and top-quark backgrounds, requirements on missing transverse energy and a $b$-jet veto are
applied. The leading lepton is required to pass the tight selection and to match the trigger, while the
subleading lepton is used as a probe. The fraction of probe leptons passing the tight criteria defines
$\epsilon_{r}$ as a function of \pt{} and $\eta$. This region is very pure in $Z$ boson events, and
comparisons of data and simulation show excellent agreement.

The fake efficiency is measured in a dijet sample selected with one baseline lepton and at least one jet with
$p_{T} \ge 35$~GeV. Requirements on the azimuthal separation between the lepton and the jet ($|\Delta\phi(\ell, j)| > 2.8$), as well as low
missing transverse momentum ($\ptmiss{} < 25$~GeV), suppress contributions from $W$+jets and $Z$+jets events with prompt
leptons. Residual prompt contributions are subtracted using simulation, and systematic uncertainties are
assigned to this subtraction. To obtain stable estimates in bins of lepton \pt{} and $\eta$, a bootstrap
resampling technique~\cite{Efron1979} is employed to determine both the central value and statistical uncertainty of
$\epsilon_{f}$.

The efficiencies are applied in an ``application region'' defined by the baseline lepton selection. The matrix method relates the observed numbers of tight and non-tight leptons to the underlying real and
fake components via $\epsilon_{r}$ and $\epsilon_{f}$. In the fake factor method, the prompt contribution in the application region is subtracted before applying the fake factor.
The resulting $j\rightarrow \ell$ yields are computed separately in the electron and muon channels and for each observable used in the measurement.

The matrix method is taken as the nominal estimate, with the difference to the fake factor method assigned as an uncertainty that is typically about 4\%. Additional systematic uncertainties are considered, including the statistical precision of the control
region measurements, the subtraction of prompt leptons in the dijet region, and differences in fake
composition between control and signal regions.
Validation regions, defined with inverted missing transverse momentum requirements, provide further
cross-checks. In all cases, the data are well described by the background model within the assigned
uncertainties.

\subsection{Electrons faking photons}

In the electromagnetic calorimeter, electrons and photons both produce narrow, localised energy deposits, making it
possible for prompt electrons to be misidentified as photons. This background, denoted $e \to \gamma$, is expected to
contribute at the ${\sim}10\%$ level to the selected data sample. This background contribution is estimated using a
data-driven fake factor method, which corrects the simulation with scale factors measured in control regions.

The basic observable is the `$e \to \gamma$ fake rate',
\[
F_{e \to \gamma} = \frac{N_{e\gamma}}{N_{ee}},
\]
defined as the ratio of reconstructed $Z \to e\gamma$ candidates, where one electron candidate is misidentified as a
photon, to reconstructed $Z \to ee$ events, where both electron candidates are identified as electrons. The $e \to \gamma$ fake rate
can be measured both in data and in the simulation, using $Z \to ee$ events. The ratio of the two defines a scale factor
\[
\mathrm{SF} = \frac{F^{\mathrm{data}}_{e \to \gamma}}{F^{\mathrm{MC}}_{e \to \gamma}},
\]
which quantifies the mismodelling of the mis-identification rate in simulation. The method is applied by
reweighting simulated events that contain an electron reconstructed as a photon with the scale
factor evaluated as a function of the transverse momentum and pseudorapidity of the misidentified photon candidate.

The event yields in the $e \to \gamma$ fake rate are measured using a tag-and-probe method in control regions enriched in $Z \to ee$ events. The leading electron candidate is required to satisfy Tight identification and isolation criteria, while the second candidate electron is treated as the probe. When the probe electron candidate is reconstructed as an electron,
the event is selected for the $ee$ control region, and when it is reconstructed as a photon, it is selected for
the $e\gamma$ control region. Fits to the invariant mass of the $ee$ or $e\gamma$ system in the range $70 < m_{\ell\ell} < 110$~GeV are
used to extract the yields of true $Z$ boson events, separating them from background contributions.
The scale factors obtained from the extracted yields are in agreement with those obtained in earlier ATLAS measurements~\cite{ATLAS:2018fzd}. The dominant systematic uncertainties arise from electron and photon identification efficiencies, energy resolution, and pile-up modelling. The $m_{\ell\ell}$ range used in the fit to extract the $Z$-boson yields is also varied. These uncertainties typically result in 5\% variations in the $e \to \gamma$ background estimate, corresponding to a 0.5\% effect on the total event yield.

\subsection{Pile-up backgrounds}

In proton--proton collisions at the LHC, multiple interactions typically occur within the same bunch crossing.
As the position of the photon production vertex is not well measured in the absence of an ID track, a selected photon can originate from an interaction that is different
from the one producing the lepton from the $W \to \ell \nu$ decay.  Such
mis-associations produce a background that mimics the $W\gamma$ signal. Although small in absolute size,
this pile-up background contribution is non-negligible.

The method to estimate the pile-up background relies on the longitudinal separation along the beam axis between the reconstructed photon
position, extrapolated to the beam line ($z_\gamma$), and the primary vertex associated with the lepton
($z_{\mathrm{PV}}$). The distribution of
\[
\Delta z = z_\gamma - z_{\mathrm{PV}}
\]
is used to distinguish genuine $W\gamma$ events from pile-up. For signal and background events where the reconstructed photon and lepton
originate from the same vertex, $\Delta z$ is sharply peaked at zero, with a width determined by the detector resolution.
In contrast, for pile-up events the two quantities are uncorrelated, giving a broader Gaussian distribution
with width ${\sim}50$~mm, consistent with the longitudinal spread of two independent proton--proton collisions.

The determination of $z_\gamma$ depends on the photon type. For converted photons, $z_\gamma$ is obtained
from tracks associated with the conversion vertex, providing sub-millimetre precision when the conversion
occurs inside the silicon detectors. For unconverted photons, $z_\gamma$ is estimated by calorimeter pointing,
using the shower profile of the photon in the first two calorimeter layers~\cite{ATLAS:2013etx}. Since the resolution of converted photons is
significantly better, the analysis uses this category to obtain a more precise estimate.

The fraction of pile-up events is extracted by comparing the observed $\Delta z$ distribution in data to the
expectation from simulated $W\gamma$ events. The signal component is normalised to the data in the central
region around $\Delta z \approx 0$, while the tails at $|\Delta z| > 50$~mm, which are dominated by pile-up, are
used to measure that contribution. This approach yields an overall pile-up fraction of
$f_{\mathrm{PU}} = (4.8 \pm 0.3)\%$. To avoid double-counting with the jet-induced photon background, which is
treated separately, a photon purity correction is applied~\cite{ATLAS:2019gey}. The resulting estimate for the pile-up photon
background is $f_{\mathrm{PU}} = (2.6 \pm 0.4)\%$. For the kinematic measurements presented in this paper,
the estimate is performed independently for each measured bin.


%
\section{Construction of CP-sensitive observables using neural networks}
\label{sec:onn}


Anomalous interactions are typically described using EFT, in which the SM Lagrangian is augmented with dimension-six operators, i.e.,
\begin{equation}
\mathcal{L} \, = \, \mathcal{L}_{\textrm{SM}} \, + \, \sum_i \frac{c_{i}}{\Lambda^2}\mathcal{O}_{i}
\label{eq:SMEFT_Lagrangian}
\end{equation}
where $\mathcal{L}_{\textrm{SM}}$ is the SM Lagrangian, the $\mathcal{O}_{i}$ are a set of the dimension-six operators, and the $c_{i}/\Lambda^2$ are Wilson coefficients that specify the strength of the anomalous interactions. The dimension-six operators can be CP-even or CP-odd. The squared scattering amplitude for the effective field theory prediction can be written as
\begin{equation}
|\mathcal{M}|^2 = |\mathcal{M_{\textrm{SM}}}|^2 \, + \, 2 \,\textrm{Re} (\mathcal{M_{\textrm{SM}}}^*\mathcal{M_{\textrm{d6}}}) \, + \, |\mathcal{M_{\textrm{d6}}}|^2 ,
\label{eq:SMEFT_ME}
\end{equation}
where $\mathcal{M_{\textrm{SM}}}$ is the SM scattering amplitude and $\mathcal{M_{\textrm{d6}}}$ is a `dimension-six' scattering amplitude that contains the anomalous interaction. The cross-section therefore contains three contributions: the SM contribution, the interference between the SM amplitude and the dimension-six amplitude, and a pure dimension-six contribution. The SM contribution and the pure dimension-six contributions are always CP-even, regardless of the nature of the operator. However, the interference contribution will be CP-even for CP-even operators and CP-odd for CP-odd operators. For CP-odd operators an equal amount of constructive interference and destructive interference occurs, meaning that the interference term integrates to zero for CP-even observables. The presence of CP-odd anomalous interactions can therefore only be probed using appropriately constructed CP-odd observables, such as \dphily{}.

The EFT samples described in Section~\ref{sec:datamc} are produced using a decomposition scheme, whereby the SM, interference and pure-dimension-six contributions are produced separately. Following the approach outlined in Refs.~\cite{Bhardwaj:2021ujv,Hall:2022bme}, an optimised CP-sensitive observable is constructed using a neural network (NN). The NN is trained to distinguish between constructive interference, destructive interference, and the SM. The outputs of the neural network are then used to construct a new observable,
\begin{equation}
\onn =  P_+  - P_- ,
\end{equation}
where $P_+$ is the probability of the event to be constructive interference and $P_-$ is the probability of the event to be destructive interference. The sum of probabilities is $P_+ + P_- + P_{SM} = 1$, where the $P_{SM}$ is the probability of the event to be a SM contribution.

The NN used in this method is a multilayer perceptron (MLP), which is composed of multiple layers of neurons, including input layer, output layer, and multiple hidden layers. All layers are arranged in a hierarchy structure and are fully connected to the subsequent layer. A dedicated MLP is trained for each CP-odd EFT operator using SM events and the related EFT interference samples. The EFT interference sample is split according to the MC event weight into two classes: positively-weighted events and negatively-weighted events, which capture the constructive and destructive interference, respectively. The MC event weight is only used to define the classes and is not used in the MLP training. The SM events make up the third class.

The MLP is constructed using Keras~\cite{chollet2015keras} and TensorFlow~\cite{Abadi:2016kic}. The Sequential model is chosen with three hidden layers of 100, 50 and 20 neurons, respectively, each employing a ReLU activation function~\cite{RELU}. Ten variables are input to the network: the transverse momentum, pseudo-rapidity, azimuthal angle and charge of the lepton, the transverse momentum, pseudo-rapidity and azimuthal angle of the photon, the magnitude and azimuthal angle of the missing transverse momentum, and the leading jet transverse momentum. To avoid inefficient training and biased learning, the input variables are scaled: the angular variables and the lepton charge are linearly scaled to the range [0,1], whereas the momentum-based variables are first log transformed and then linearly scaled to the range [0,1]. The scaling parameters for each variable are determined using the training dataset. There are three output classifications, SM, constructive interference, and destructive interference. The output layer therefore has three neurons and uses softmax activation functions. The categorical cross-entropy~\cite{Shannon:1948nms} loss function is chosen (with one-hot encoding) alongside the Adam optimiser~\cite{Kingma:2014vow} for the network weights and biases. The batch size is chosen to be 5000 and the networks are trained for up to 200 epochs. Early stopping is implemented with a min-delta of zero and a patience of 3 or 10, for the \oHWBtil{} and \oWtil{} operators, respectively. The min-delta and patience choices were chosen to avoid the training being stopped too early, a situation that can arise due to the kinematic similarity between the constructive- and destructive- interference classes.

For both classifiers, the input data is split into training data (80\%), validation data (10\%) and test data (10\%). The training data is used to train the MLPs and the validation data is used in the early stopping criteria. The test data is used to test for overtraining, with no obvious biases observed in the reconstructed \onn{} distributions. Each MLP achieves an accuracy of around 56\%, which is reasonable compared to random chance (33\%). Receiver-operating characteristics are checked for each class in the model using the test data and the area-under-the-curve found to be between 0.71 and 0.81 depending on the class and model.

Figure~\ref{fig:dataMCcomparison_onn} shows the number of events measured in the baseline region as a function of \onn{} for the \oHWBtil{} and \oWtil{} operators. The data are in good agreement with the combined prediction containing the signal and all background processes. The signal and prompt backgrounds are modelled using the MC simulations discussed in Sec.~\ref{sec:datamc}. The other backgrounds are determined using the data-driven methods discussed in Sec.~\ref{sec:backgrounds}. A data augmentation approach is taken when deriving the data-driven backgrounds for \onn{} (and \dphily{}): the distributions are filled twice using values of \onn{} and \onn$_{,\mathrm{CP}}$, where \onn$_{,\mathrm{CP}}$ is the observable obtained after applying a CP-transform to the data. A weight of 0.5 is used for each event when filling the histograms twice to preserve the overall normalisation. The data augmentation ensures that the data-driven backgrounds are CP-even in the observable and improves the statistical accuracy of the data-driven background by a factor of $\sqrt{2}$.

\begin{figure}[tp]
\subfloat[]{\includegraphics[width=0.5\textwidth]{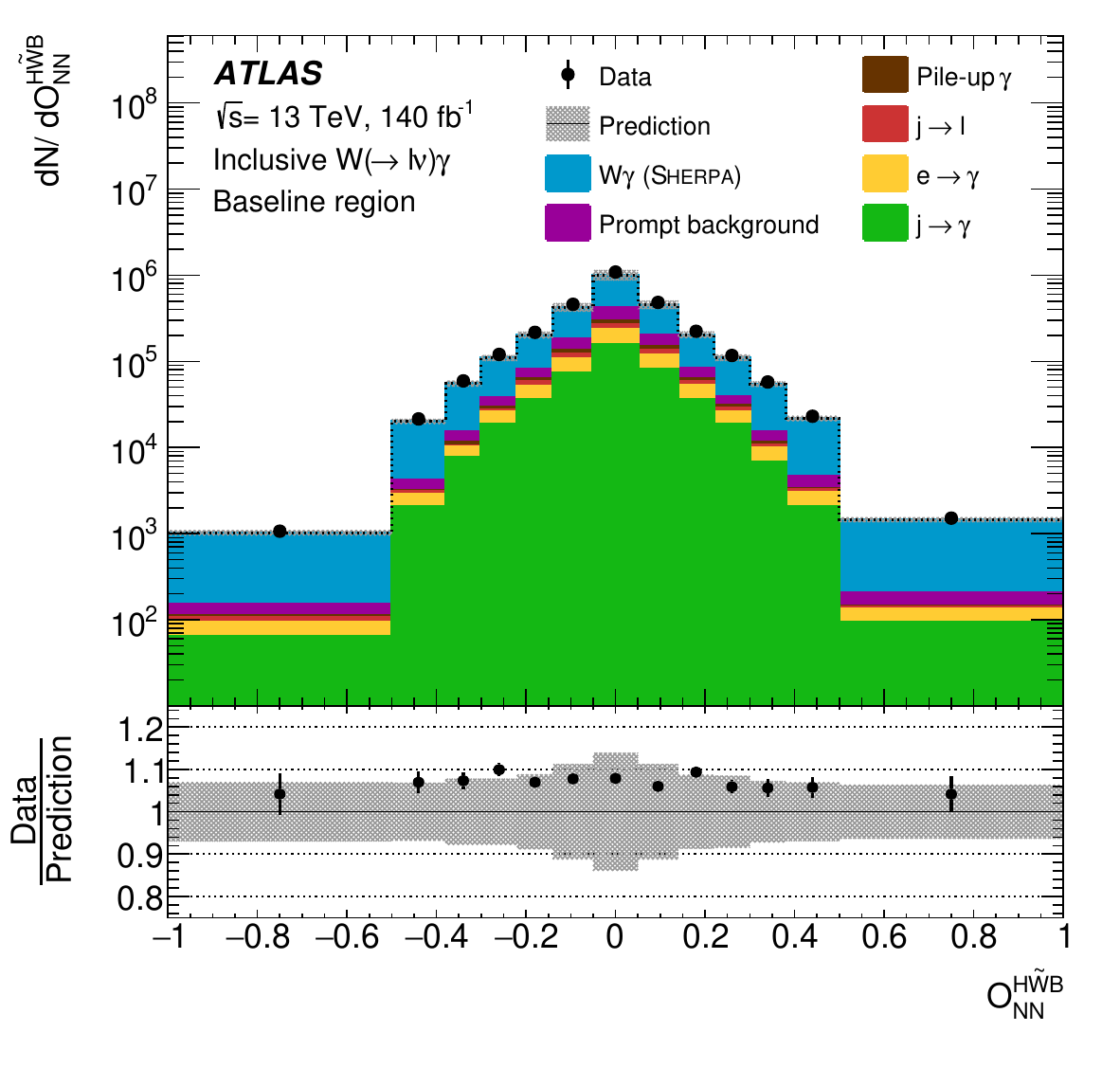}}
\subfloat[]{ \includegraphics[width=0.5\textwidth]{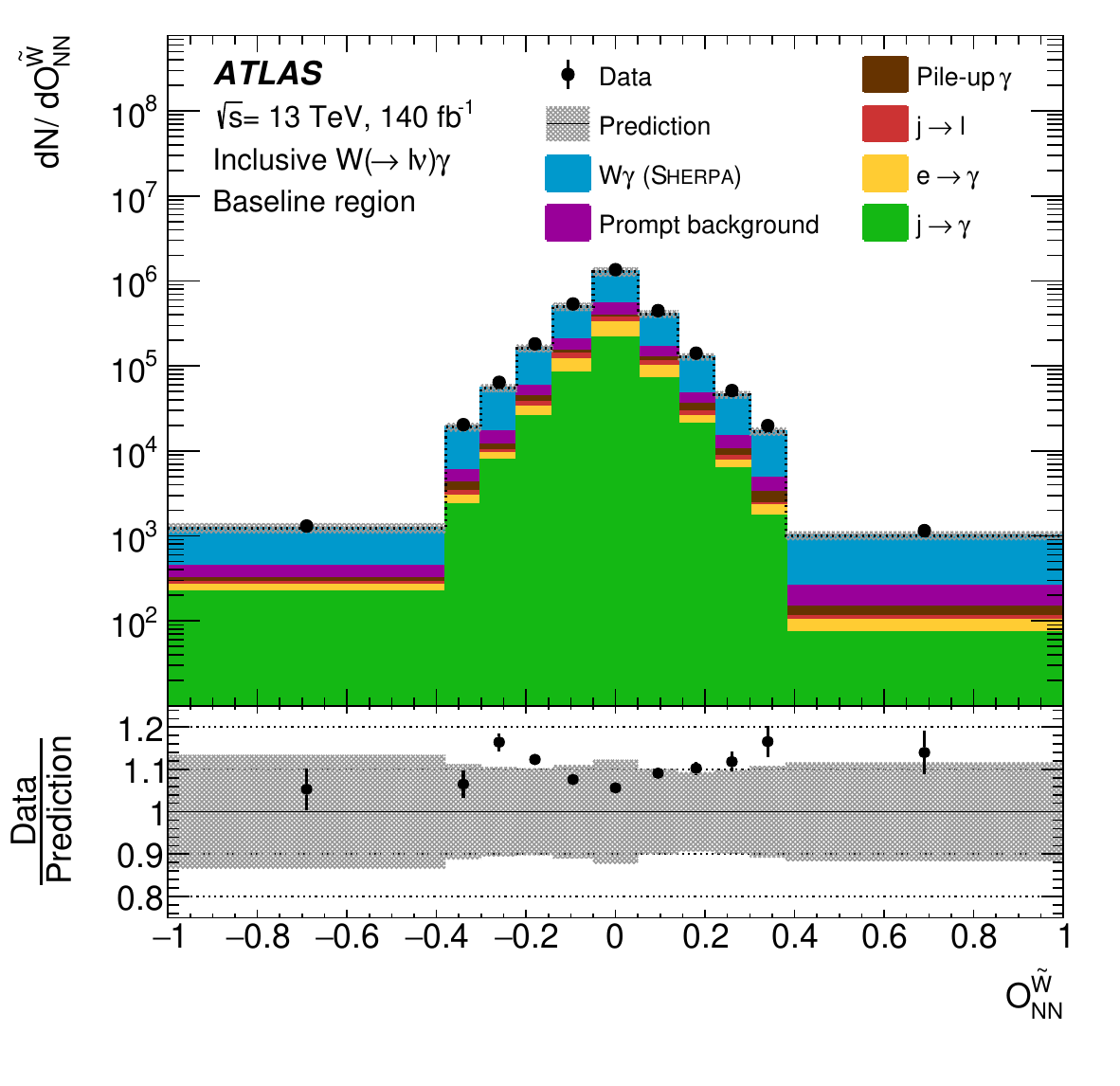}}
\caption{Predicted and observed yields as a function of \onn{} for (a) the \oHWBtil{} operator and for (b) the \oWtil{} operator. The data are represented as black points and the associated error bars represent the statistical uncertainty. Background processes with prompt leptons and photons are estimated by using simulations and labelled as `Prompt'.   Backgrounds arising from non-prompt leptons ($j\rightarrow \ell$), non-prompt photons ($j\rightarrow \gamma$), electrons-faking-photons ($e\rightarrow \gamma$) and pile-up photons are estimated using data-driven techniques. The total uncertainty on the combined signal and background prediction is shown as a grey band.
\label{fig:dataMCcomparison_onn}}
\end{figure}


%
\section{Correction for detector effects}
\label{sec:unfolding}

Differential cross sections for inclusive $W\gamma$ production are measured at the particle level, by subtracting the backgrounds from the observed event yields in each bin of each distribution and then correcting for the effects of detector inefficiency and resolution.

The particle level is defined using objects and selections that closely resemble the detector-level selections presented in Section~\ref{sec:evsel}. Final state particles are defined as those with $c\tau>10$~mm. Prompt particles are defined as those that do not originate from the decay of a hadron. Charged leptons are defined by dressing prompt electrons and prompt muons with all nearby prompt photons, i.e. those within a distance $\Delta R <0.1$. Leptons from $\tau$ decays are ignored. Dressed electrons are required to have $\pt>7~$GeV and $|\eta|<2.47$, but excluding the region $1.37<|\eta|<1.52$. Dressed muons are required to have $\pt>7$~GeV and $|\eta|<2.5$. Photons are required to be prompt and satisfy $\pt>12$~GeV and $|\eta|<2.37$, but excluding the region $1.37<|\eta|<1.52$. Jets are defined using the anti-$k_t$ algorithm with radius parameter of $R=0.4$ and are required to have $\pt>20$~GeV and $|y|<4.4$. The input to the jet finding algorithm are all final state particles excluding prompt leptons. Jets are identified as $b$-jets if they contain a $B$-hadron with $\pt>5$~GeV within $\Delta R<0.3$ of the jet centre. The missing transverse momentum is defined as the vector sum of all final-state non-interacting prompt particles. An overlap removal procedure is then applied to the reconstructed leptons, photons and jets to mimic the procedure applied at detector level. Jets are removed if they are within $\Delta R<0.4$ of a photon or $\Delta R<0.2$ of a lepton. Following this, photons and jets are removed if they are within $\Delta R<0.4$ of a lepton.

The \textit{baseline} fiducial region for the majority of the differential measurements is defined as containing exactly one lepton, at least one photon, zero $b$-jets, and $\ptmiss>40$~GeV. The lepton and the leading photon are required to have $\pt>30$~GeV. Photons are also required to be isolated, i.e. the scalar sum of transverse momenta of all interacting particles within $\Delta R<0.4$ of the photon is required to satisfy $p_{\textrm T}^\textrm{iso} < 0.026\pty + 8.21$~GeV. The photon and lepton are required to be separated by $\Delta R>0.8$ and have an invariant mass of $\mly<81$~GeV or $\mly>101$~GeV. For the differential cross sections measured as a function of \dphily{} and \detaly{}, events are also required to have no jets with $\pt>30$~GeV; this is referred to as the \textit{jet veto} fiducial region. The double differential cross section as a function of \pty{} and $\phi_{f}$ is measured in both the \textit{baseline} and \textit{jet veto} regions. Finally, the measurements of the boost asymmetry are unfolded to a common phase space for leptons, i.e. $|\eta_\ell|<2.5$.

Each distribution is unfolded to particle level using D'Agostini's iterative unfolding method \cite{DAgostini:1994fjx}. This method uses Monte Carlo simulations to (i) remove events that are reconstructed at detector level but not in the fiducial region at particle level (fiducial fraction), (ii) reverse the migrations between bins of distributions, and (iii) correct for events that are in the fiducial region at particle level but not reconstructed at detector level (reconstruction efficiency). The \sherpa{} $W\gamma$ sample is used as the default simulation for unfolding, with the EW $W\gamma \! j \! j$ contribution included. The binning of each distribution is chosen such that the purity of each bin is typically greater than 60\% and that the expected statistical accuracy of the background-subtracted data is around 1-2\%. The number of iterations is chosen to minimise the overall uncertainty, accounting for statistical uncertainties and bias in the unfolding method. The unfolding bias is determined using a data-driven method and described in Sec.~\ref{sec:systematics}. Two iterations are chosen for all distributions except \onnHWBtil{} and \onnWtil, where three iterations are needed.

The unfolding procedure introduces statistical correlations between neighbouring bins in the measured distributions. These correlations are small (1-4\%) for the \pty{}, \ptl{}, \etay{}, \detaly{}, \dphily{} and \mly{} distributions, but rise to 10\% for the \mtlvy{} distribution. The largest correlations (around 20\%) are observed for some bins in the \onnWtil{} and \onnHWBtil distributions, and for the double differential distributions involving the $W$ boson decay angles.


%
\section{Systematic uncertainties}
\label{sec:systematics}

Experimental systematic uncertainties arise from lepton reconstruction, photon reconstruction, jet reconstruction, $b$-jet identification, missing transverse momentum reconstruction, the pile-up of multiple proton--proton interactions, and the luminosity of the dataset.

The lepton trigger efficiencies and lepton and photon reconstruction, identification and isolation efficiencies in simulation are corrected on an event-by-event basis using scale factors as outlined in Sec.~\ref{sec:datamc}. Systematic uncertainties associated with this procedure are determined by varying the scale factors by their corresponding uncertainties~\cite{EGAM-2018-01,EGAM-2021-01,EGAM-2021-02,MUON-2018-03,MUON-2022-01}. Lepton and photon energy and momentum scales are also calibrated using data and the  calibrations are accounted for by scaling and smearing the lepton and photon transverse momenta accordingly. The overall impact of the muon uncertainties and the electron efficiency uncertainties on the measured differential cross sections is typically less than 1\%. The impact of photon efficiency uncertainties is larger, being about 2.5\% at low \pty{} and 1.5\% at high \pty. The impact of electron/photon energy scale and resolution systematics rises from sub-percent at low \pty{} to 3\% at high \pty{}. For the boost asymmetry measurements, the lepton and photon uncertainties impact the numerator and denominator of the ratio in a correlated way and largely cancel.

The jet energy scale is calibrated using \pt{}- and $\eta$- dependent correction factors as described in Sec.~\ref{sec:evsel}. Systematic uncertainties on these correction factors are propagated through the analysis by scaling and smearing the reconstructed jet momentum accordingly~\cite{JETM-2018-05,JETM-2022-06}, which accounts for residual differences between data and simulation in the jet energy scale and resolution. Uncertainties due to the inefficiency of the jet vertex tagger~\cite{PERF-2014-03} and $b$-jet identification~\cite{FTAG-2018-01,FTAG-2019-07} algorithms are evaluated by applying scale factors to the MC event weights. The impact of jet energy scale and resolution uncertainties on the measured differential cross sections are typically around 2\%, except in the jet veto phase space and at low \mtlvy{}, where the impact is 5\% and 6\%, respectively. The impact of flavour-tagging uncertainties is less than 0.5\%. Again, these experimental uncertainties largely cancel for the boost asymmetry measurements.

The missing transverse momentum is evaluated using the negative vector sum of all identified electrons, muons, photons and jets, as well as a soft-term to account for low-\pt{} activity. The uncertainties in the leptons, photons and jets reconstruction and calibration are propagated directly to the missing transverse momentum calculation. An additional uncertainty associated with the energy scale and resolution of the soft term is evaluated independently~\cite{JETM-2020-03} and found to have an impact of about 1\% -- 1.5\% on the differential cross sections.

The uncertainty in the luminosity at ATLAS is determined to be 0.83\% for the Run~2 dataset~\cite{DAPR-2021-01}, using a combination of Van Der Meer beams scans and measurements taken by the LUCID-2 detector~\cite{LUCID2}. The pile-up modelling in simulation involves overlaying the simulated hard-scatter events with inelastic proton--proton collisions as described in Sec.~\ref{sec:datamc}. The uncertainty in the pile-up modelling is obtained by reweighting the distributions to account for residual differences between the data and simulation. The impact of this uncertainty is typically less than 1\%, but as high as 1.5\% at very low \ptl{}.

Prompt backgrounds are impacted by all of the experimental uncertainties discussed above. Theoretical uncertainties in the estimation of the prompt backgrounds are obtained by independently varying the renormalisation scale, $\mu_{\textrm R}$, and factorisation scale, $\mu_{\textrm F}$, by factors of 0.5 and 2.0, with the added requirement that  $0.5<\mu_{\textrm R} / \mu_{\textrm F}<2.0$. This produces six possible variations, each of which changes the normalisation and shape of the background. These variations are propagated through the analysis. The final uncertainty in the differential cross section is taken to be the envelope of the results obtained from the six variations. PDF uncertainties are accounted for using the quadrature sum of the deviations induced by the 100 eigenvectors of the Hessian NNPDF set.
The uncertainty in the PDFs due to the choice of $\alpha_S$ is estimated separately by NNPDF by varying the value of $\alpha_S$ by $\pm$0.001; this is added in quadrature with the uncertainty obtained from the eigenvectors. The combined impact of scale and PDF variations in the prompt backgrounds is found to be typically 2\%. Statistical uncertainties in the prompt backgrounds are propagated through the analysis directly and found to be typically 0.2\%.

The uncertainties in the data-driven background estimates for $j\rightarrow \gamma$, $j\rightarrow \ell$, $e\rightarrow \gamma$, and pile-up (PU) photons, are described in Sec.~\ref{sec:backgrounds}. These uncertainties are propagated directly through the analysis chain. The statistical uncertainties in these estimates arise from data and are included in the statistical uncertainty of the unfolded measurement. The largest systematic uncertainty is associated with the $j\rightarrow \gamma$ estimate, which is around 4\% at low \pty{} but falling to 1\% at high \pty{}. The systematic uncertainty associated with the $j\rightarrow \ell$ estimate is typically about 1.5\% but rises to 4\% for low \ptl{}. The impacts of the systematic uncertainties in the PU photon estimate and in the $e\rightarrow\gamma$ estimate are negligible.

The unfolding procedure is outlined in Sec.~\ref{sec:unfolding} and uses inclusive $W\gamma$ simulations produced with \sherpa{} to correct the background-subtracted event yields for the effects of detector inefficiency and resolution. The signal simulations are affected by all of the experimental uncertainties discussed above. In addition, theoretical uncertainties in the signal modelling are evaluated in four ways. First, the factorisation scales and renormalisation scales in the default \sherpa{} simulation are varied by factors of 0.5 and 2.0, with the restriction that   $0.5<\mu_{\textrm R} / \mu_{\textrm F}<2.0$. Second, the uncertainty due to PDF choice is evaluated in the same way as for the prompt backgrounds, using the 100 eigenvector variations of the NNPDF set and by varying the choice of $\alpha_S$ used in the PDFs. Third, a data-driven test of the bias in the unfolding is estimated by independently reweighting the response matrix such that the \sherpa{} simulation better matches the data for each measured distribution. Finally, the impact of the choice of event generator is determined by using the \mgpy{} simulation of $W\gamma$ events, but after reweighting the \mgpy{} response matrix to match the \sherpa{} response matrix to avoid double counting the uncertainty estimated using the data-driven bias test.
The latter is the largest of the four contributions, which are treated as independent and uncorrelated estimates of the signal modelling uncertainty.
The overall impact of the signal modelling uncertainties on the differential cross sections is typically 2-3\%, but can rise to over 10\%  at high $|\onn{}|$ and at high transverse momentum in the $\pty\times\phi_f$ distributions.

The impact of each source of uncertainty on the differential cross sections is shown in Figure~\ref{fig:sys_breakdown} for the \pty{}, \etay{}, \detaly{} and \onn{} distributions. The systematics associated with the data-driven background are typically the dominant uncertainties in the measurements, with signal modelling uncertainties being important at high-\pty{} and high-$|\onn|$.

\begin{figure}[tp]
\subfloat[]{\includegraphics[width=0.5\textwidth]{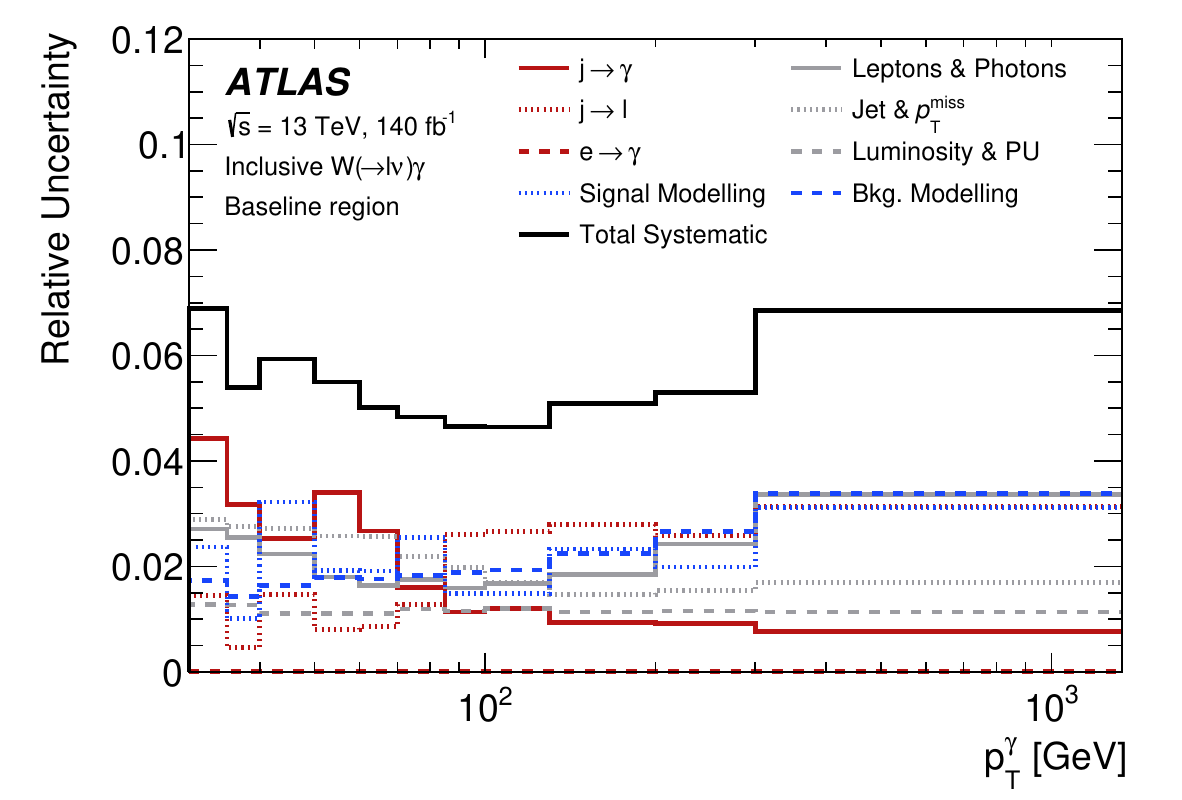}}
\subfloat[]{\includegraphics[width=0.5\textwidth]{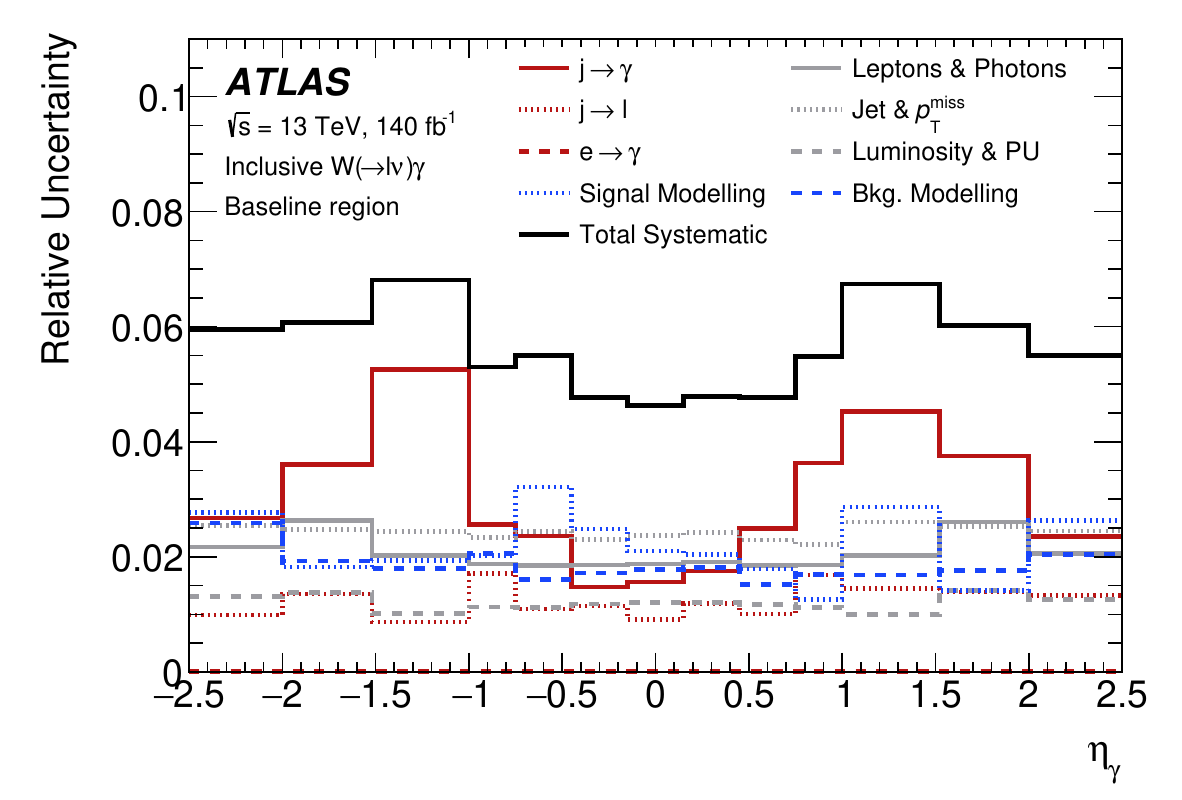}}\\
\subfloat[]{\includegraphics[width=0.5\textwidth]{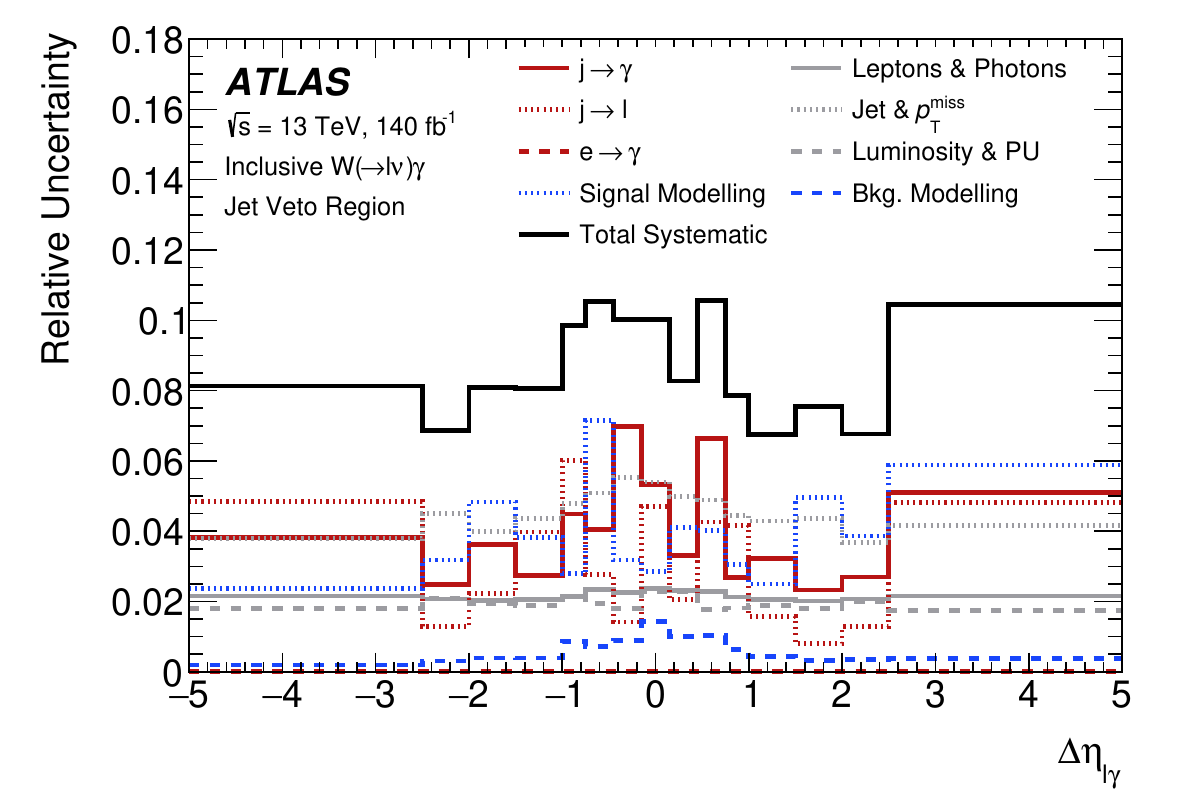}}
\subfloat[]{\includegraphics[width=0.5\textwidth]{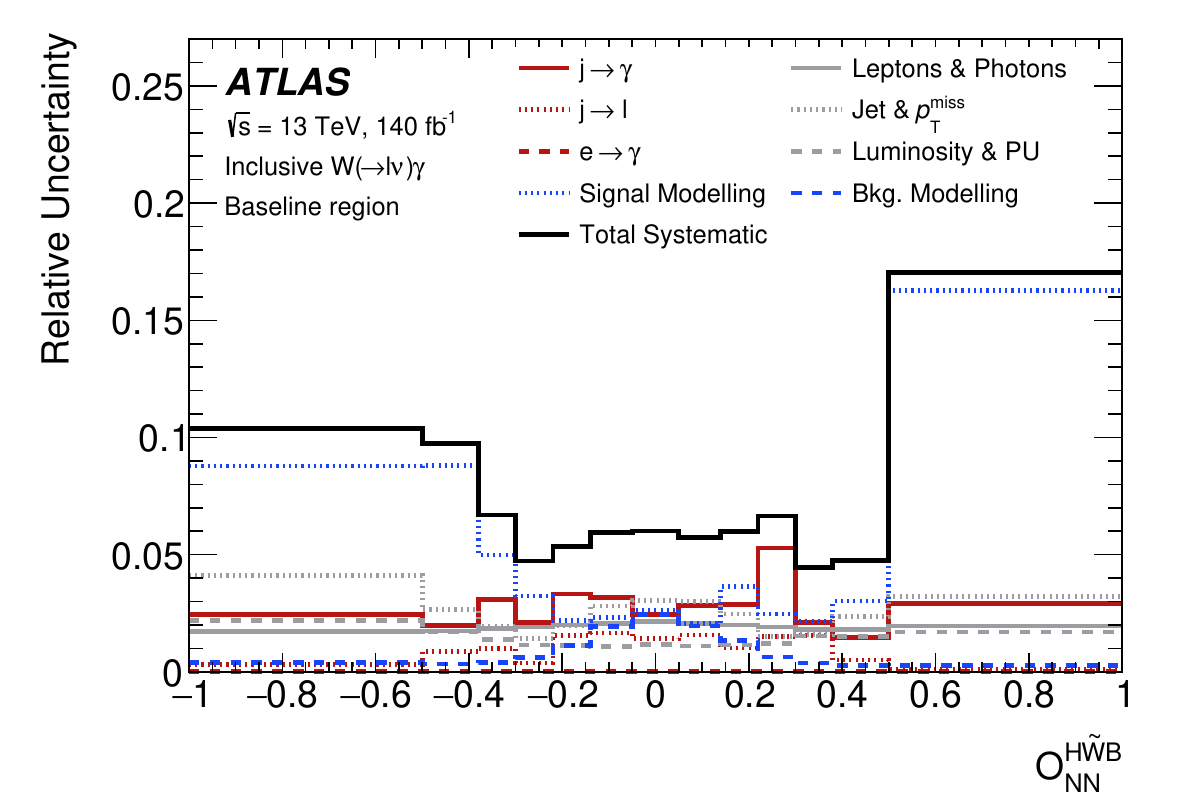}}
\caption{Relative uncertainties on the differential cross-sections as a function of (a) \pty{}, (b) \etay{}, (c) \detaly{} and (d) \onn{}. Sources of uncertainty are grouped together for clarity. The total systematic uncertainty is shown as the quadrature sum of each source of uncertainty.
\label{fig:sys_breakdown}}
\end{figure}


%
\section{Results}
\label{sec:result}

In this section, the detector-corrected measurements that probe the kinematic properties of the $W\gamma$ system are presented. This includes differential cross sections sensitive to the radiation amplitude zero effect, the polarisation of the $W$ boson, and the CP-structure of the $WW\gamma$ triple gauge interaction, as well as boost asymmetry observables that can be used in global PDF fits.

\begin{figure}[thbp]
\centering
\subfloat[]{\includegraphics[width=0.49\textwidth]{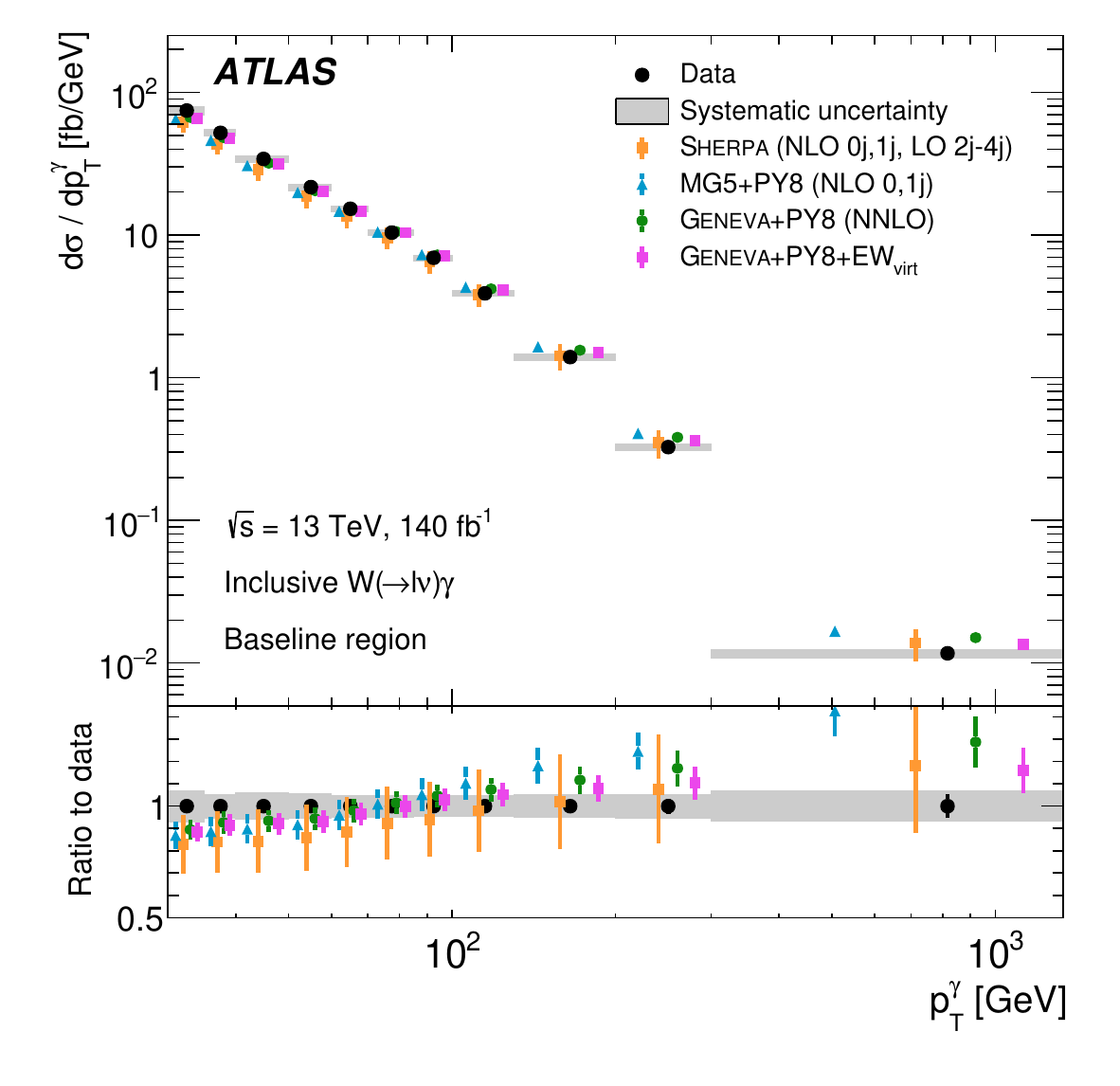}}
\subfloat[]{\includegraphics[width=0.49\textwidth]{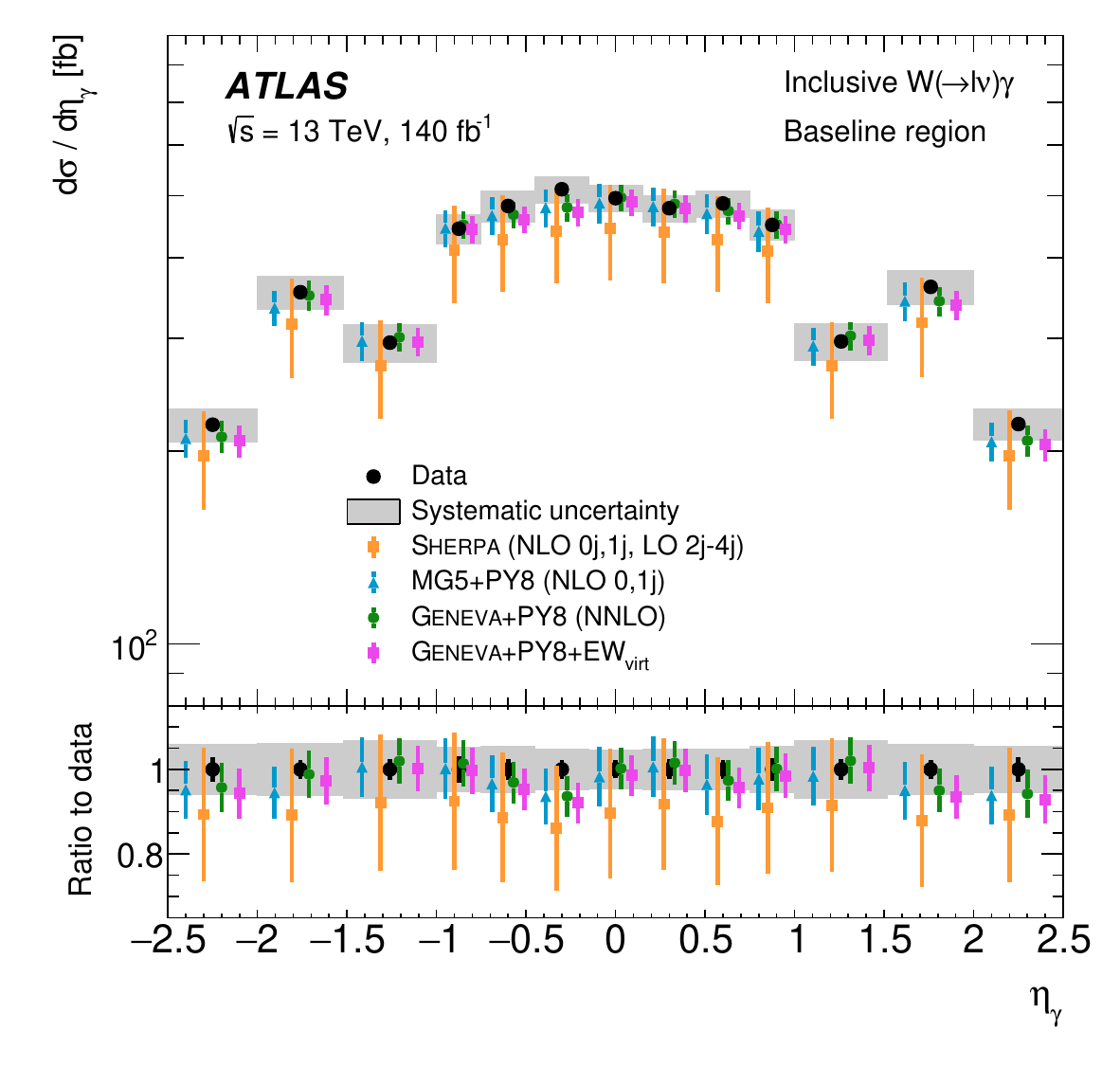}}\\
\subfloat[]{\includegraphics[width=0.49\textwidth]{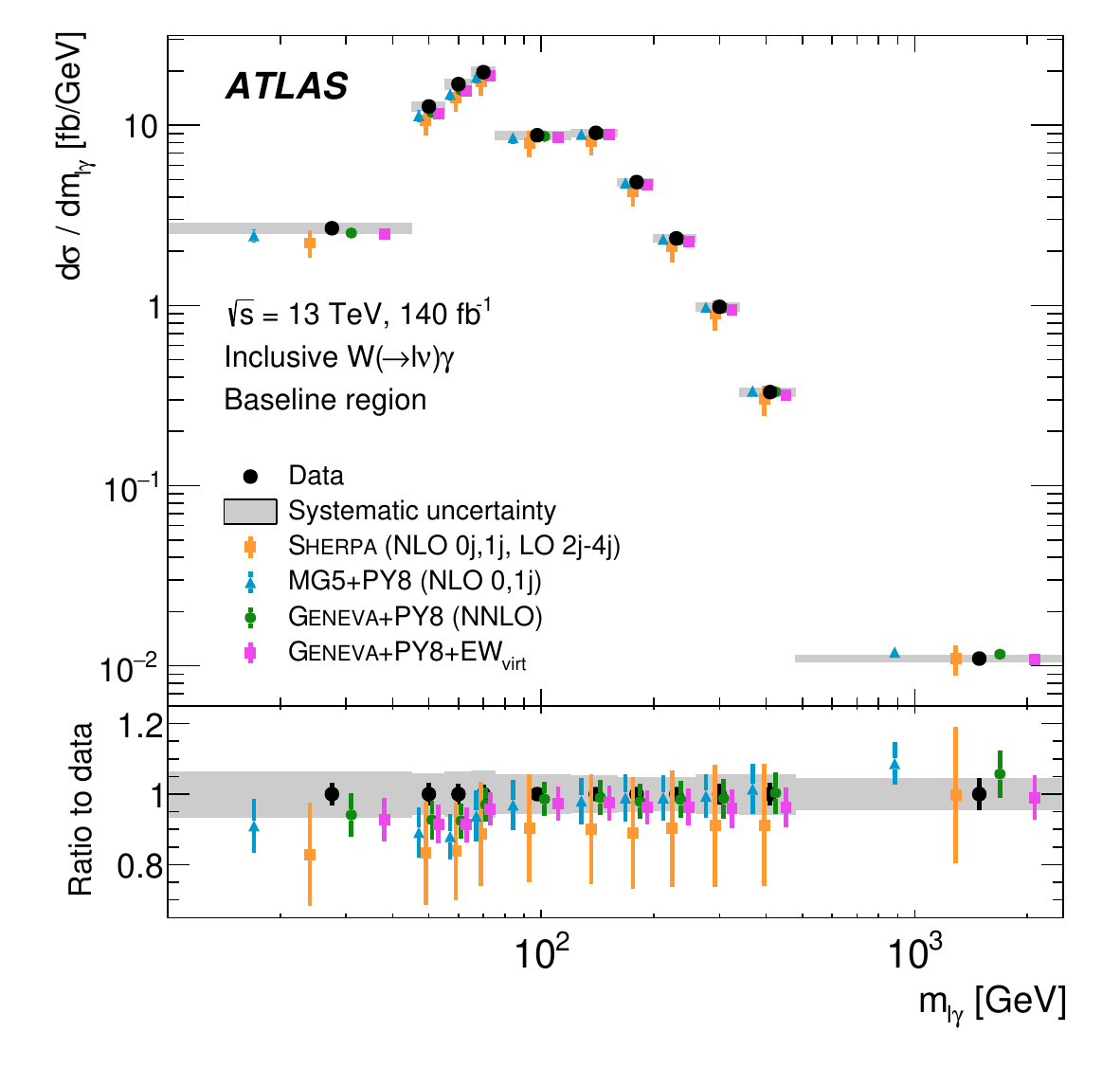}}
\subfloat[]{\includegraphics[width=0.49\textwidth]{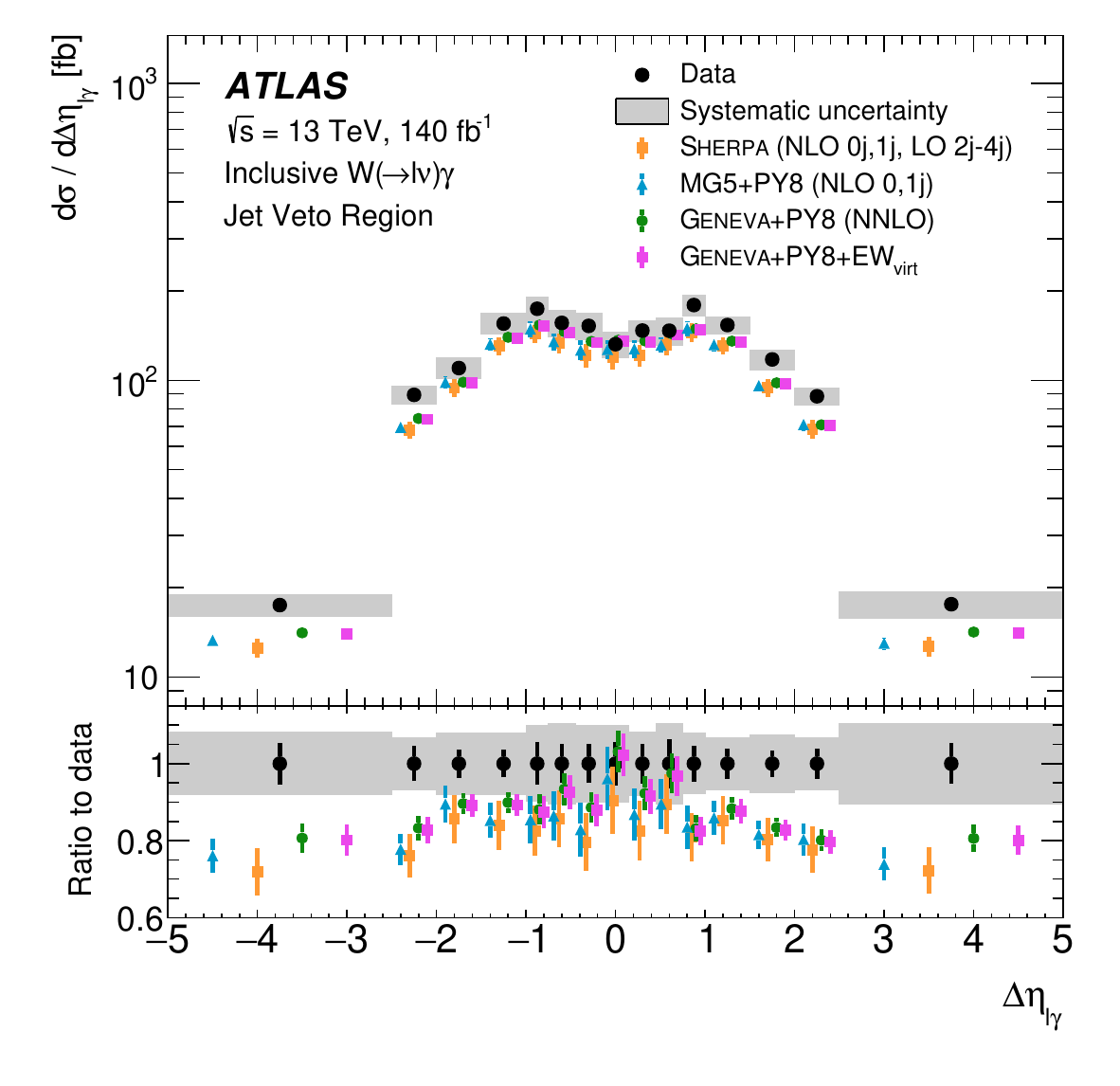}}
\caption{Differential cross sections measured as a function of
(a) \pty{}, (b) \etay{}, (c) \mly{} and (d) \detaly{}. The unfolded data are shown as black points, with the statistical uncertainty represented by an error bar and the systematic uncertainty represented as a grey band. The data are compared with theoretical predictions produced by \sherpa{} (orange points), \mgpy{} (blue points) and \genpy{} (green points). The \sherpa{} and \mgpy{} predictions are accurate to NLO in QCD for zero and one jets. \genpy{} is accurate to NNLO for inclusive $W\gamma$ production. The \best{} prediction (pink points) is accurate to NNLO in QCD and NLO for virtual EW corrections. Uncertainty bands derived from scale variations are shown for all four theoretical predictions. Each theory prediction is slightly offset from the bin centre to avoid overlap.}
\label{fig:diff_xs1}
\end{figure}

The differential cross sections are compared to four different Standard Model particle-level predictions for inclusive $W\gamma$ production. Two of these are constructed from the \sherpa{} and \mgpy{} samples presented in Section~\ref{sec:datamc}. \sherpa{} is accurate to NLO in QCD for zero and one jets in the final state and to LO for up to four additional jets. \mgpy{} is accurate to NLO in QCD for zero and one jets in the final state. \geneva{}~\cite{Alioli:2012fc,Alioli:2013hqa,Alioli:2015toa,Cridge:2021hfr,Alioli_hepmc} is used to provide an inclusive $W\gamma$ prediction that is accurate to NNLO in QCD. \geneva{} matches a resummed soft collinear effective field theory prediction to fixed order NNLO predictions and is interfaced to \pythia{} for parton showering, hadronisation and underlying event. The \NNPDF[3.1nnlo] PDF is used and the CP5 set of tuned parameters~\cite{CMS:2022awf}. This prediction is referred to as \genpy{}. A final prediction is produced by scaling the \genpy{} prediction with virtual EW corrections estimated with \sherpa{}. This prediction is referred to as \best{} in the subsequent discussion. Uncertainties on the \sherpa{} and \mgpy{} predictions are determined by independently varying the renormalisation and factorisation scales by factors of 0.5 and 2.0 and taking the envelope of all such variations. Uncertainties on the \genpy{} prediction arise from renormalisation/factorisation scale variations as well as variations in the resummation scales (beam scale, soft scale, and profile transition). The total uncertainty on \genpy{} is taken to be the quadrature sum of two envelopes, one formed from the renormalisation and factorisation scale variations and one formed from the resummation scale variations. The uncertainty on \genpy{} is propagated to the \best{} prediction.

Figures~\ref{fig:diff_xs1} (a)-(c) show the differential cross section measurements as a function of \pty{}, \etay{}, and \mly{} in the \textit{baseline} fiducial region. The \sherpa{} prediction is in agreement with the data for all three distributions but has large theoretical uncertainties. The \mgpy{} prediction has smaller uncertainties than \sherpa{} and is in agreement with the data at low \pty{} but it overestimates the cross section at high-\pty{}. The \genpy{} prediction has smaller uncertainties than both \sherpa{} and \mgpy{} but also overestimates the cross section at high-\pty{}. The \best{} prediction has the best agreement with the data across all kinematic regions, with the NLO EW virtual corrections playing a crucial role at high momentum scales.

Figure~\ref{fig:diff_xs1} (d) shows the differential cross section as a function of \detaly{} in the \textit{jet veto} fiducial region. This variable can be used to probe the radiation amplitude zero (RAZ) effect, which is predicted at LO in QCD but is removed when higher-order QCD corrections are included in the calculation. Imposing the jet veto restricts real emissions and enhances the RAZ effect. The data and the theoretical predictions demonstrate the dip at $|\detaly| = 0$ that is a characteristic of RAZ. Although all theoretical predictions are below the data at high values of $|\detaly|$, it is possible that their uncertainties are underestimated due to the application of a jet veto when measuring a process that has large higher-order corrections~\cite{Stewart:2011cf}.

Figure~\ref{fig:polarisation_observables1} (a) shows the double differential cross section as a function of
$\theta_f$ and $\phi_f$, which are the polar and azimuthal angles of the negative helicity fermion from the $W$ boson decay as reconstructed in the special reference frame defined in Section~\ref{sec:intro}. This distribution is measured in the \textit{baseline} fiducial region. These angular variables probe the polarisation state of the $W$ boson and provide direct sensitivity to the $W$ boson spin-density matrix. The measured angular distributions are in good agreement with the Standard Model predictions, indicating the data are consistent with the expected $W$ boson polarisation states for inclusive $W\gamma$ production.

\begin{figure}[tbph]
\centering
\subfloat[]{\includegraphics[width=0.99\textwidth]{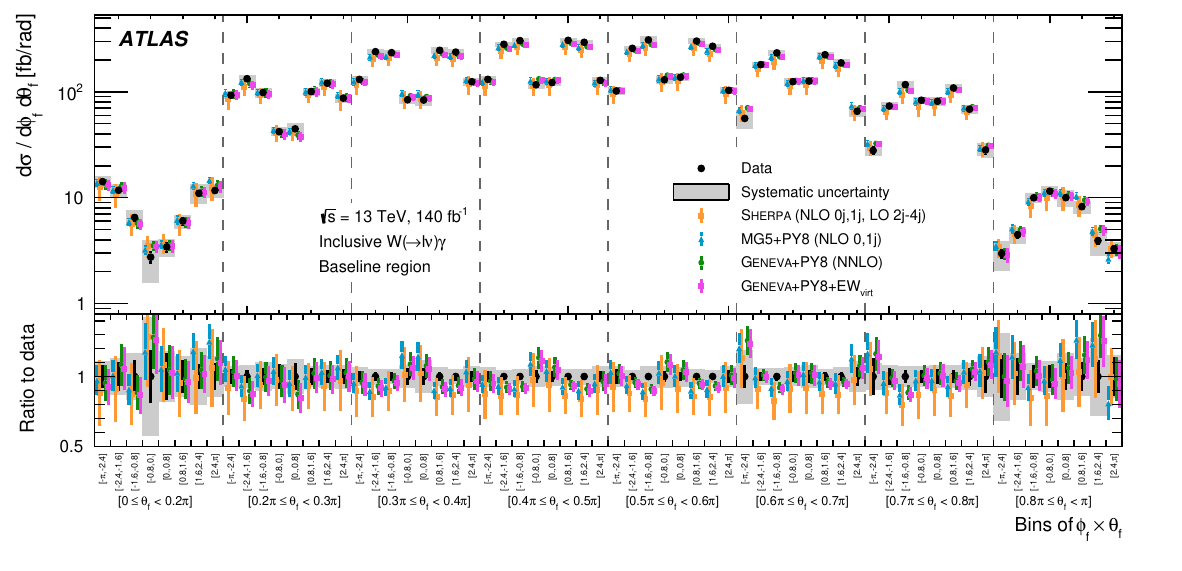}}\\
\subfloat[]{\includegraphics[width=0.99\textwidth]{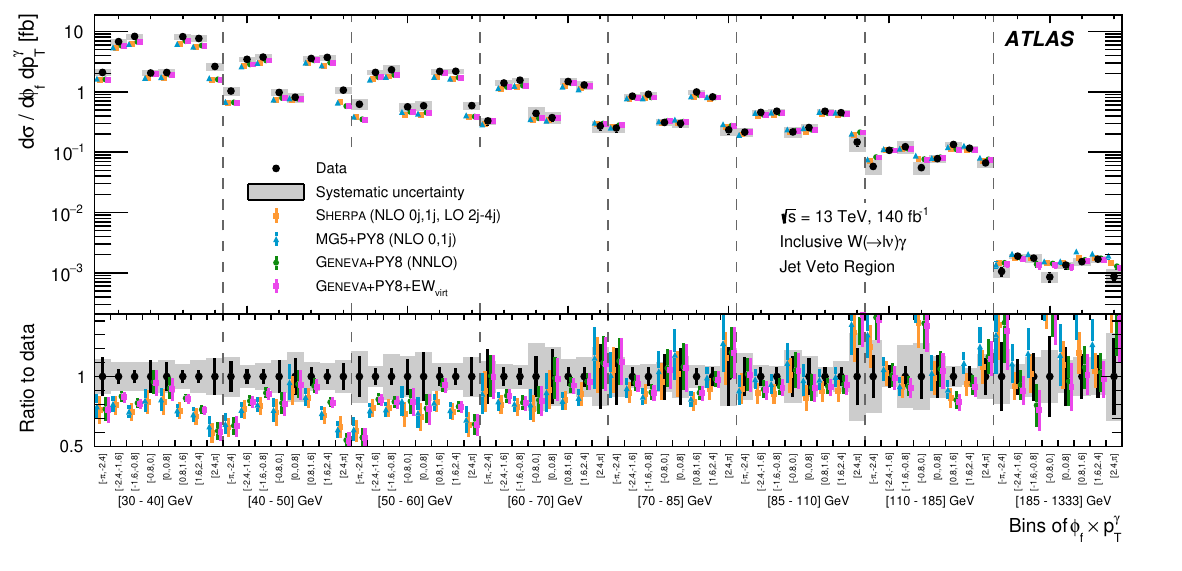}}
\caption{(a) Double differential cross section as a function of $\phi_f$ and $\theta_f$. (b) Double differential cross section as a function of $\phi_f$ and \pty{}. The vertical dashed lines highlight the bins of $\theta_f$ in the top figure and the bins of \pty{} in the bottom figure. %
The data and theoretical predictions are presented in the same way as for Figure~\ref{fig:diff_xs1}.}
\label{fig:polarisation_observables1}
\end{figure}

Figure~\ref{fig:polarisation_observables1} (b) shows the double differential cross sections as a function
of $\phi_f$ and \pty{}. The $\phi_f$ observable is sensitive to the interference between scattering amplitudes that contain $W$ bosons in different polarisation states. The $\phi_f$ observable was proposed in Ref.~\cite{Panico:2017frx} as a way to probe EFT interference effects for CP-even operators. Measuring the double differential distribution as a function of $\phi_f$ and \pty{} then increases these interference effects because contributions from dimension-six EFT operators are typically enhanced with respect to the SM at high momentum scales. The measurement is performed in the \textit{jet veto} fiducial region, which additionally suppresses the SM with respect to the EFT contributions due to the RAZ effect. The data are again in good agreement with the SM predictions, except at lower  \pty{} values. This discrepancy may be related to the use of a jet veto, where scale variations may not provide a reliable estimate of the theoretical uncertainty.

Figure~\ref{fig:cp_observables} shows the differential cross section as a function of the optimised observables, \onn{}, which are constructed from the outputs of a
multiclass neural network (described in Section~\ref{sec:onn}). The \onn{} observables allow the interference effects predicted by anomalous CP-violating $WW\gamma$ interactions to be probed. The measured distributions are in good agreement with the Standard Model expectations.

\begin{figure}[tbph]
\centering
\subfloat[]{\includegraphics[width=0.49\textwidth]{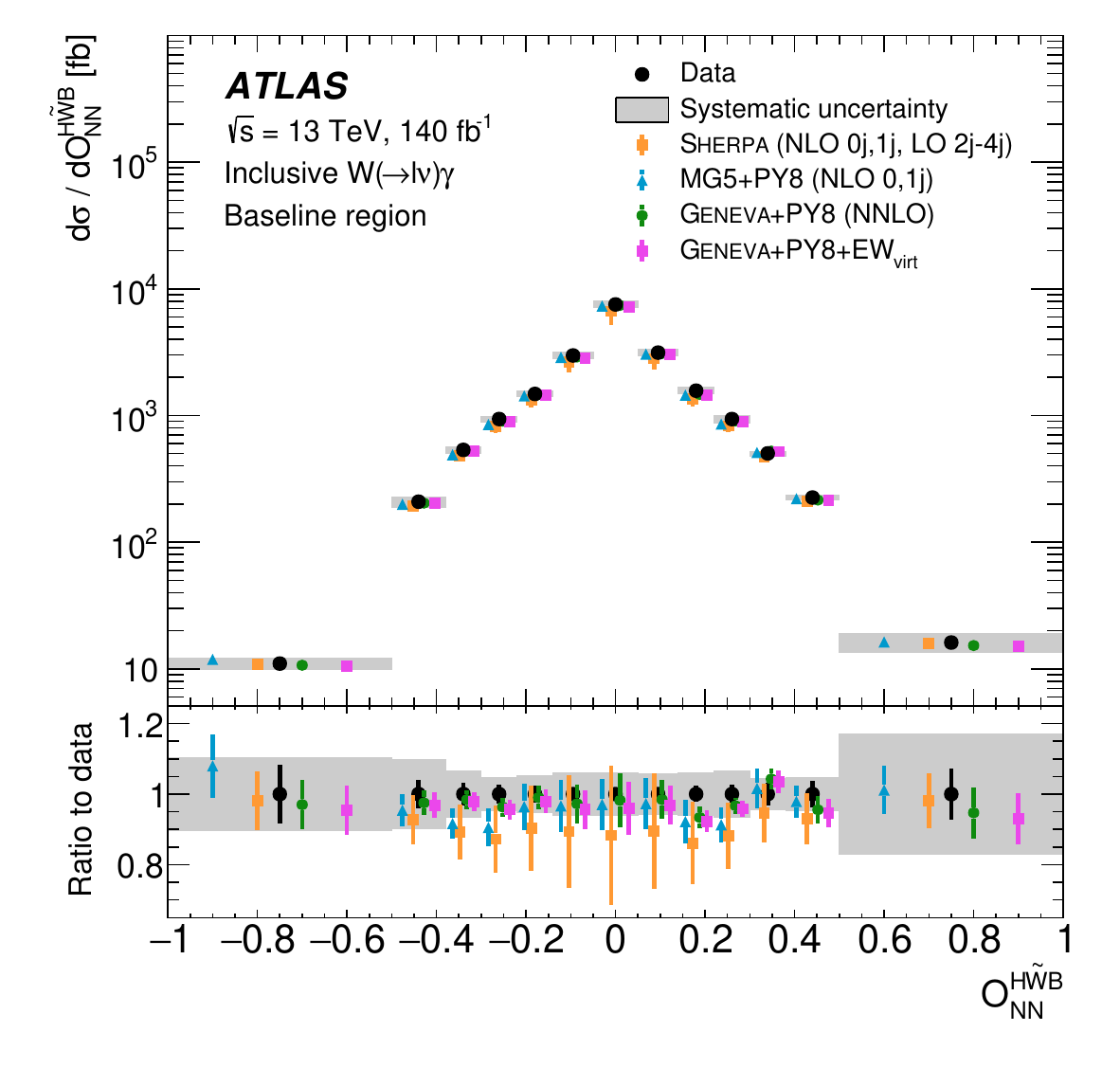}}
\subfloat[]{\includegraphics[width=0.49\textwidth]{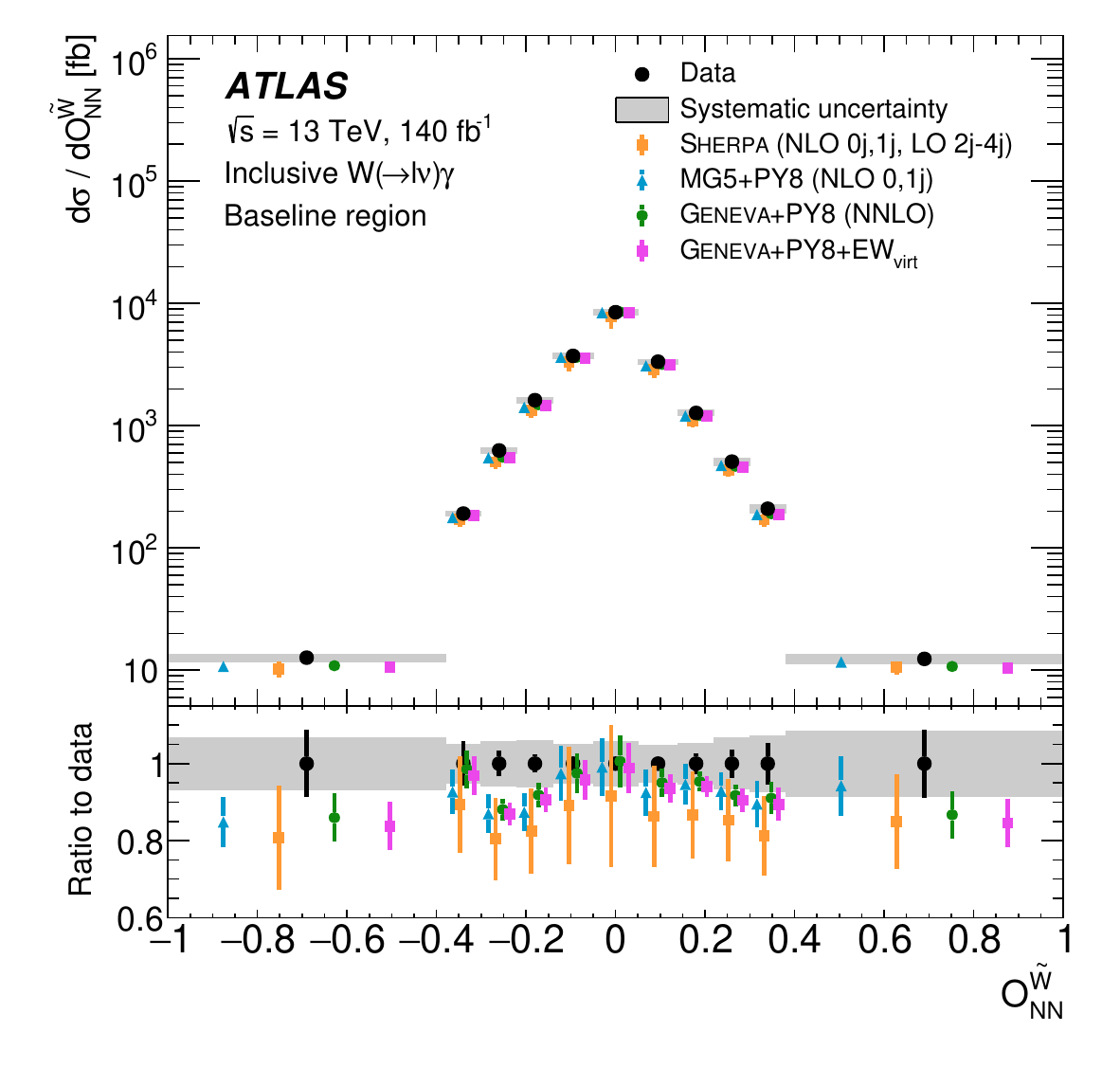}}
\caption{Differential cross sections as a function of (a) \onnHWBtil{} and (b) \onnWtil{}, which are CP-sensitive observables constructed from the outputs of neural networks, as discussed in Section~\ref{sec:onn}. The data and theoretical predictions are presented in the same way as for Figure~\ref{fig:diff_xs1}.}
\label{fig:cp_observables}
\end{figure}

\begin{figure}[tbp]
\centering
\subfloat[]{\includegraphics[width=0.49\textwidth]{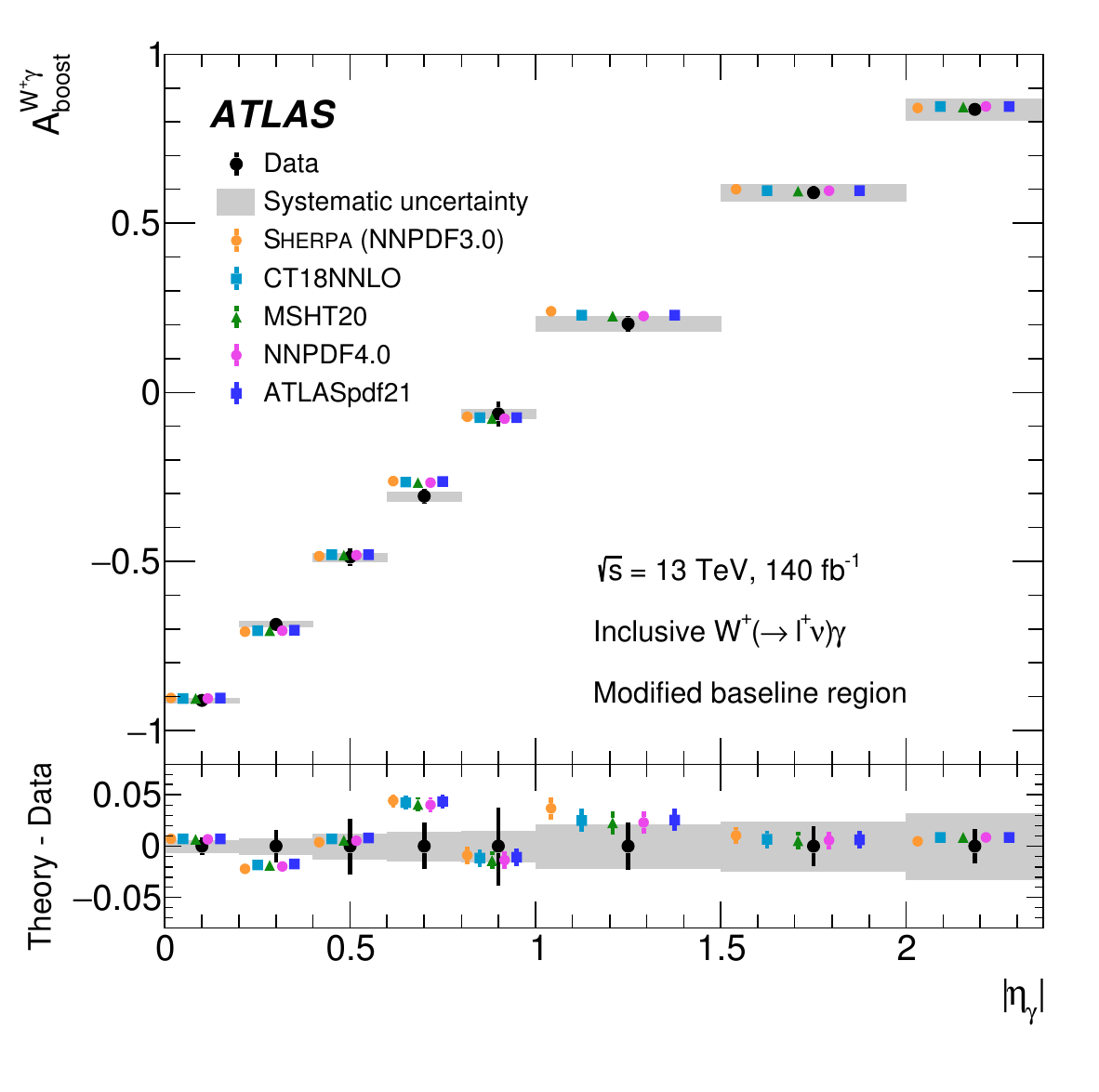}}
\subfloat[]{\includegraphics[width=0.49\textwidth]{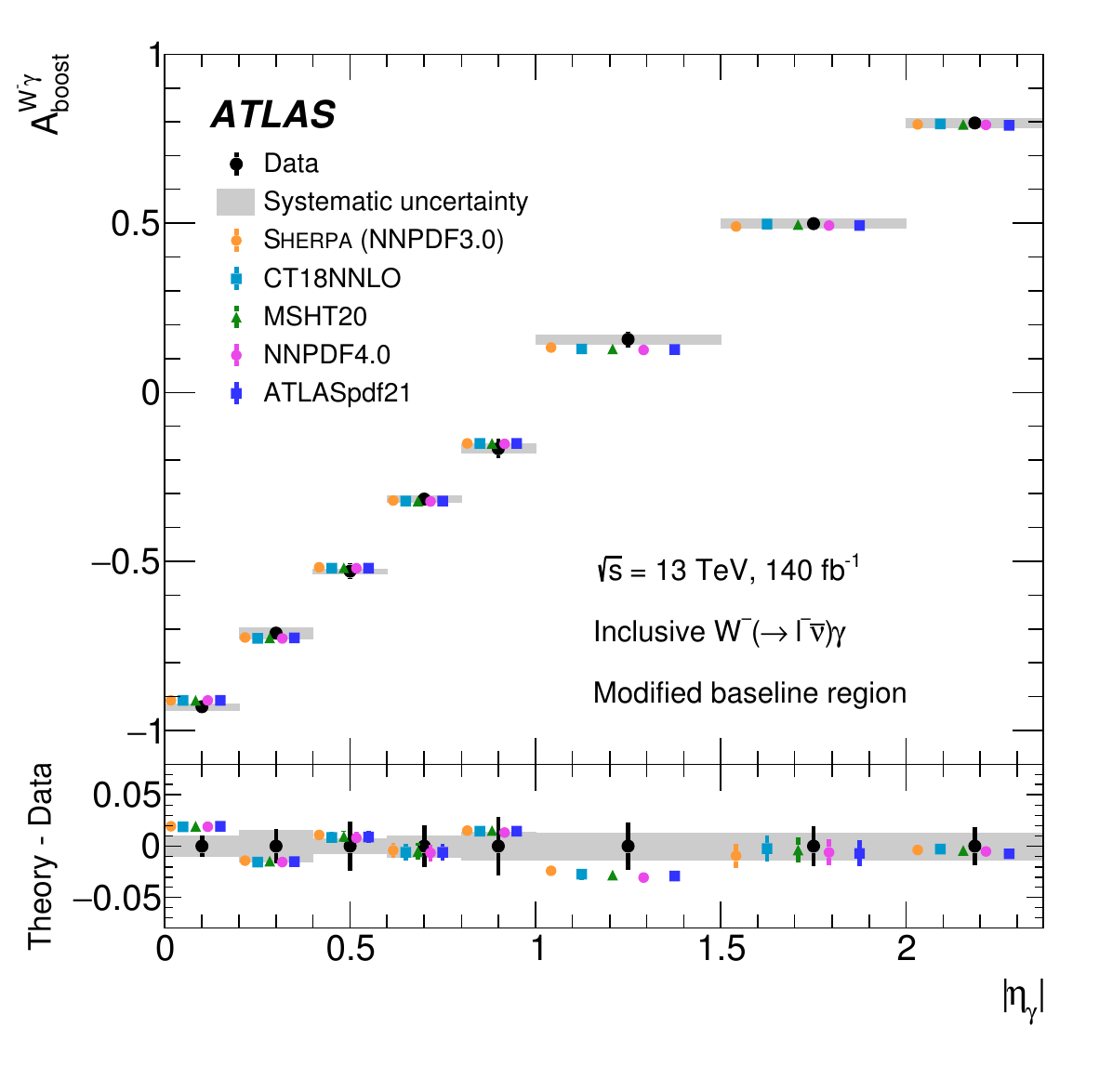}}
\caption{Differential measurements of the boost asymmetry as a function of $|\eta_{\gamma}|$ for (a)
for $W^{+}\gamma$ production and for (b) $W^{-}\gamma$ production. The data are presented in the same way as for Figure~\ref{fig:diff_xs1} and compared to theoretical predictions obtained with \sherpa{} but using different PDF sets. The uncertainty on the theoretical predictions include renormalisation/factorisation scale variations and PDF uncertainties. }
\label{fig:boost_asymmetry}
\end{figure}

Additional differential cross sections, measured as a function of \ptl{}, \mtlvy{} and \dphily{}, are presented in the Appendix. The \ptl{} and \mtlvy{} distributions are measured in the \textrm{baseline} regions, whereas the \dphily{} measurement is performed in the \textit{jet veto} region. In addition, a double differential cross section as a function of $\phi_f$ and \pty{} in the \textit{baseline} fiducial region is also presented. The agreement between the data and the theoretical predictions is similar to the other observables shown in this section.

PDF-sensitive observables are studied using the $W\gamma$ boost asymmetry,
$A^{W\gamma}_{\mathrm{boost}}$, defined in Section~\ref{sec:intro}. The detector corrected differential cross sections that are used in the boost asymmetry calculation are measured separately for events containing leptons and antileptons, and also separated by lepton flavour. They are unfolded to the modified \textit{baseline} fiducial region discussed in Section~\ref{sec:unfolding} that uses a common selection for lepton/antilepton pseudorapidity. The electron and muon channel measurements are then combined to reduce the statistical uncertainty using the BLUE method~\cite{Nisius:2020jmf}. The measured boost asymmetries are
\begin{equation}
A_\text{boost}^{W^-\gamma} %
= -0.091 \pm 0.012\text{(stat.)} \pm 0.017\text{(sys.)} \pm 0.002\text{(th.)},
\label{equ:CombBoostAsymmetry}
\end{equation}
\begin{equation}
A_\text{boost}^{W^+\gamma} %
= 0.033 \pm 0.012\text{(stat.)} \pm 0.009\text{(sys.)} \pm 0.003\text{(th.)}.
\label{equ:CombBoostAsymmetry2}
\end{equation}
These are in good agreement with the predicted values of $-0.101 \pm 0.003$ for the $W^-\gamma$ process and $0.043 \pm 0.003$ for the $W^+\gamma$ process, obtained using \sherpa{} with the \NNPDF[3.0nnlo] PDF set.

Differential measurements of the boost asymmetry as a function of $|\eta_{\gamma}|$ are shown in Figure~\ref{fig:boost_asymmetry} for both $W^{+}\gamma$ and $W^{-}\gamma$ production. The data are compared to \sherpa{} predictions produced with different PDF sets~\cite{Hou:2019efy,Bailey:2020ooq,NNPDF:2021njg,STDM-2020-32}.  Boost asymmetries contain complementary information that can help untangle the contributions of valence quarks and sea quarks \cite{Yang:2022bxv}.
The measurements are well described by the Standard Model predictions for all PDF sets considered.
Additional differential measurements of the boost asymmetry as a function of $|\eta_{\ell}|$ are presented in the Appendix.


%
\section{EFT interpretation}
\label{sec:eft}


The differential cross section measurements for inclusive $W\gamma$ production can be interpreted in terms of physics beyond the Standard Model, using the framework of the Standard Model Effective Field Theory (SMEFT), introduced in Section~\ref{sec:onn}. For $W\gamma$ production, the focus is on the operators \oHWB, \oW, \oHWBtil{} and \oWtil{}, which modify the weak-boson self-interactions and the interaction between the weak bosons and the Higgs boson.

Each operator in the SMEFT ($\mathcal{O}$) has an associated Wilson coefficient ($c/\Lambda^2$) that describes the strength of the anomalous interactions. The scattering amplitude squared contains three contributions: the SM contribution, an interference term proportional to $c/\Lambda^2$, and a pure dimension-six contribution proportional to $c^2/\Lambda^4$, as shown in Equation~\ref{eq:SMEFT_ME}.
The SMEFT prediction in bin $i$ of a differential distribution is therefore parametrised as
\begin{equation}
\sigma_i^{\text{SMEFT}} = \sigma_i^{\text{SM}} + \frac{c}{\Lambda^2} \, \sigma^\textrm{int} + \frac{c^2}{\Lambda^4} \, \sigma^\textrm{d6}
\end{equation}
where $\sigma_i^{\text{SM}}$ is the SM cross section
and $\sigma^\textrm{int}$ and $\sigma^\textrm{d6}$ are the interference and pure dimension-six contributions, respectively, evaluated at a Wilson coefficient of unity. The interference and pure dimension-six contributions are determined from the SMEFT simulated samples presented in Sec.~\ref{sec:datamc}.

Figure~\ref{fig:eft_2d} shows the deviations from the SM prediction induced by each operator as a function of \pty{} and $\phi_f$. The deviations are presented at representative values of the Wilson coefficients and broken down into the interference contribution, the pure dimension-six contribution, and both contributions combined. The interference terms for CP-even (CP-odd) operators induce symmetric (antisymmetric) deformations relative to the SM as a function of $\phi_f$. Furthermore, as expected the EFT effects typically increase with \pty{} relative to the SM. Figure~\ref{fig:eft_cpodd} shows the deviation from the SM predictions induced by the CP-odd operators as a function of \dphily{} and \onn{}. The interference (pure dimension-six) contributions are asymmetric (symmetric). The interference contribution is larger, relative to the SM, for the \onn{} distributions than for the \dphily{} distribution, indicating a better separation of the constructive and destructive interference.

\begin{figure}[thbp]
\centering
\includegraphics[width=0.8\textwidth]{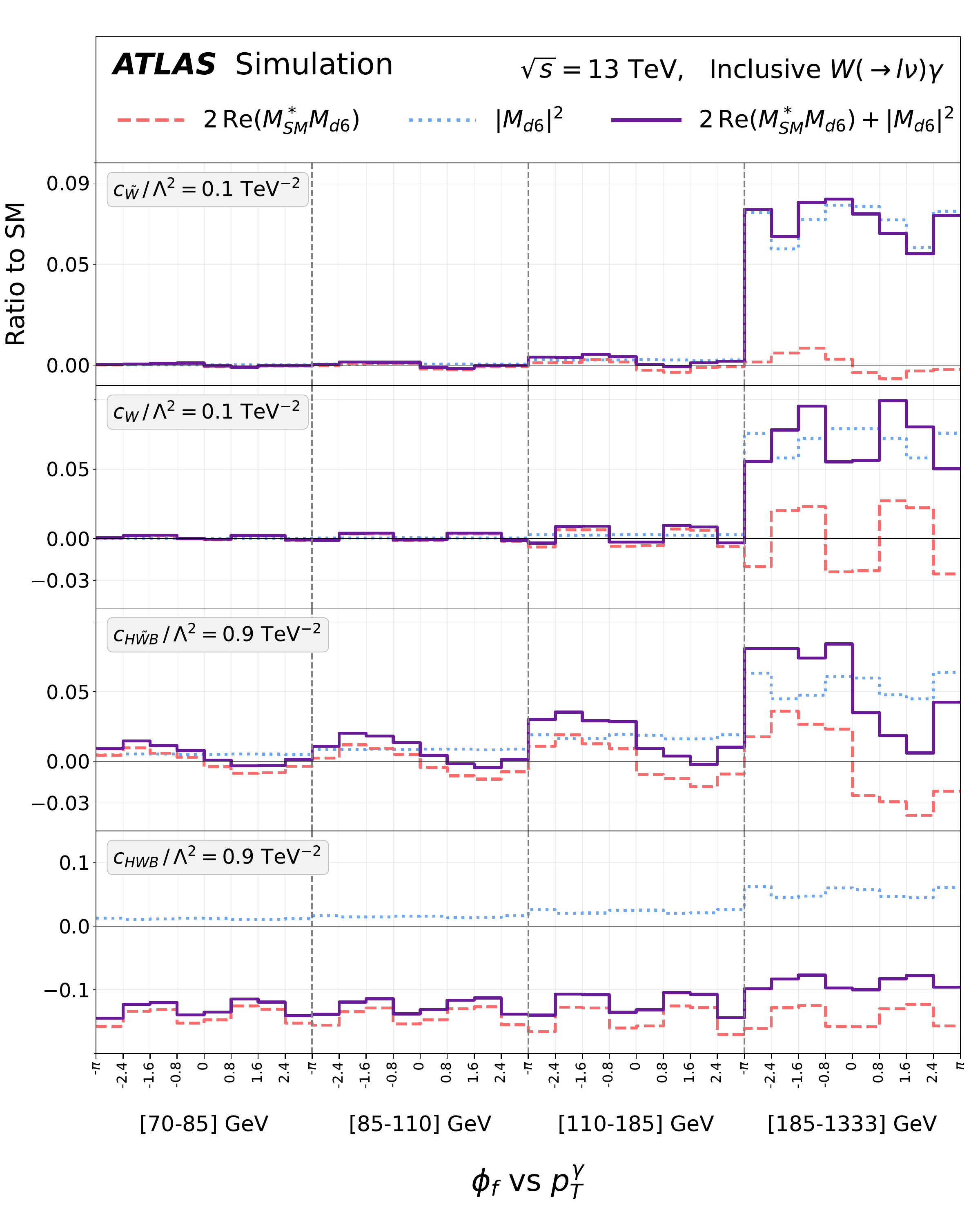}\\
\caption{Impact of the \oW{}, \oHWB{},\oWtil{} and \oHWBtil{} operators on the inclusive $W\gamma$ double differential cross section measured as a function of \pty{} and $\phi_f$. The deviations are shown for a subset of the measured bins at high-\pty{}. The expected contributions from the pure dimension-six term ($|\mathcal{M_{\textrm{d6}}}|^2$) and from the interference between the SM and dimension-six amplitudes ($2 \,\textrm{Re} (\mathcal{M_{\textrm{SM}}^{*}}\mathcal{M_{\textrm{d6}}})$) are shown relative to the pure-SM prediction and represented as dotted and dashed lines, respectively. The total contribution to the $W\gamma$ cross-section is shown as a solid line.}
\label{fig:eft_2d}
\end{figure}

\begin{figure}[thbp]
\centering
\includegraphics[width=0.8\textwidth]{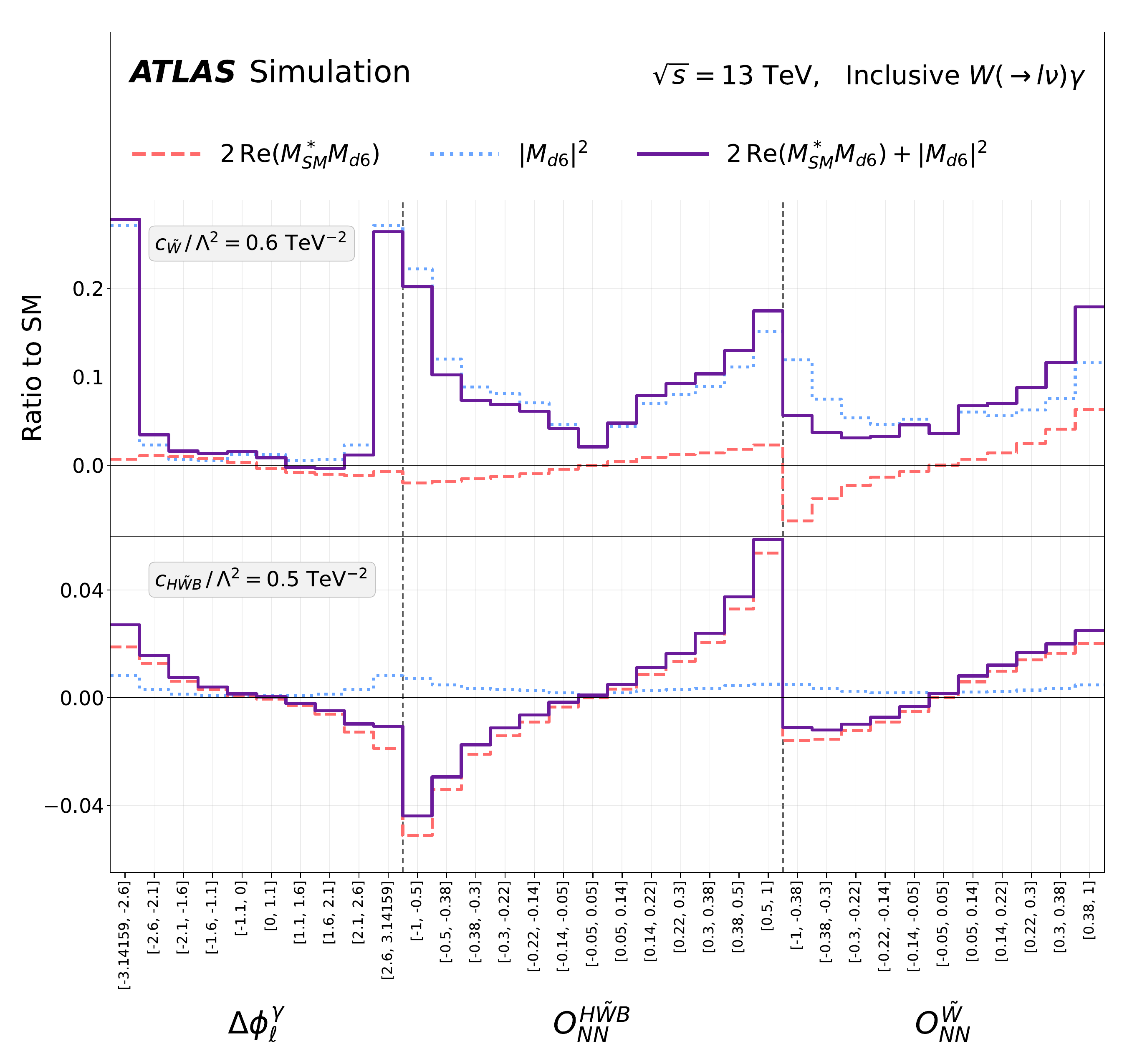}\\
\caption{Impact of the \oW{}, \oHWB{},\oWtil{} and \oHWBtil{} operators on the inclusive $W\gamma$  differential cross section measured as a function of \dphily{}, \onnWtil{} and \onnHWBtil{}. The theoretical predictions are presented in the same way as for Fig.~\ref{fig:eft_2d}.}
\label{fig:eft_cpodd}
\end{figure}

The CP-odd operators, \oHWBtil{} and \oWtil{}, are constrained using the differential cross section measurements as a function of the optimised observables based on neural network discriminants, \onnHWBtil{} and \onnWtil{}, respectively. Constraints on the CP-even operators, \oHWB{} and \oW{}, are extracted in the \textit{baseline} and \textit{jet veto} regions, respectively, using the double-differential cross section measurements as a function of \pty{} and $\phi_f$. The most sensitive distribution for each operator was evaluated using Asimov data constructed from the \sherpa{} simulation. The \onn{} observables improve the expected limits on the CP-odd operators by a factor of nearly two compared to the use of \dphily{} alone.

The SMEFT simulations are produced at LO accuracy in QCD for zero and one jets in the final state, and are therefore well below the accuracy of the SM predictions presented in this paper.
$k$-factors can be applied to the interference and pure-dimension-six contributions to recover the higher-order QCD corrections, under the assumption that they factorise from the EFT corrections. The $k$-factors are determined bin-by-bin using the SM prediction, i.e. $k=\sigma_{\sherpa}^{\textrm{SM}}/\sigma_\textrm{EFT}^{c_j=0}$, where $\sigma_{\sherpa}^{\textrm{SM}}$ is the SM prediction evaluated using \sherpa{} and $\sigma_\textrm{EFT}^{c_j=0}$ is the prediction obtained from the SMEFT framework with all Wilson coefficients set to zero. However, in the case of $W\gamma$ production, the $k$-factors calculated from the SM samples are about 1.5-2. For this reason, the constraints on the Wilson coefficients are presented with and without the $k$ factor applied. The factorisation of QCD corrections and EFT corrections has been investigated using the \textsc{Smeft@nlo} model~\cite{Degrande:2020evl}, which provides EFT predictions accurate to NLO in QCD, but only for CP-even operators. The NLO QCD $k$-factor was found to similar for both $c_W/\Lambda^2=0$ and $c_W/\Lambda^2=1$~TeV$^{-2}$ and have similar kinematic dependences.

The unfolded differential cross sections and the EFT-dependent theoretical predictions are used to define a likelihood function. Statistical correlations between the bins of the measured distributions are evaluated during the unfolding procedure and included in the likelihood using a covariance matrix. Experimental systematic uncertainties are included as Gaussian-constrained nuisance parameters and each source is treated as fully correlated across the bins of the observable but uncorrelated to other sources. Uncertainties in the theoretical prediction due to renormalisation and factorisation scale uncertainties are also implemented as Gaussian-constrained nuisance parameters, with two nuisance parameters used to account for uncertainties in the normalisation and shape of the theoretical prediction. For observables measured in the \textit{baseline} region, the standard envelope of the scale variations is used. For observables measured in the \textit{jet veto} region, the theoretical uncertainties are evaluated with the Stewart-Tackmann method~\cite{Stewart:2011cf}, which provides a conservative treatment of scale variations in exclusive jet bins. Confidence levels are calculated using the profile-likelihood test statistic, which is assumed to follow a $\chi^2$ distribution with one degree of freedom \cite{Wilks:1938dza}. This allows the 95\% confidence intervals to be constructed for each Wilson coefficient.

\begin{table}[tbp]
\caption{Expected and observed 95\% confidence-level (CL) intervals on selected dimension-six
Wilson coefficients derived from the $W\gamma$ differential cross-section
measurements.
For each coefficient, limits are obtained using the observable providing the
strongest sensitivity and are shown for all combinations of inclusion of the pure
dimension-six contribution ($|\mathcal{M}_{d6}|^{2}$) and application of a $k$-factor.}
\label{tab:eft_limits}
\centering
\begin{tabular}{cccccc}
\toprule
\multirow{2}{*}{%
\begin{tabular}{c}
Wilson \\
coefficient
\end{tabular}}
& \multirow{2}{*}{Observable}
& \multirow{2}{*}{Includes $|\mathcal{M}_{d6}|^2$}
& \multirow{2}{*}{$k$-factor}
& \multicolumn{2}{c}{95\% CL [$\mathrm{TeV}^{-2}$]} \\
\cmidrule(lr){5-6}
& & & & Expected & Observed \\
\midrule
\multirow{4}{*}{$c_W/\Lambda^2$}
& \multirow{4}{*}{$\phi_f \times \pty$}
& no  & no  & $[-0.37, 0.38]$ & $[-0.42, 0.36]$ \\
& & yes & no  & $[-0.12, 0.11]$ & $[-0.06, 0.06]$ \\
& & no  & yes & $[-0.23, 0.25]$ & $[-0.28, 0.21]$ \\
& & yes & yes & $[-0.09, 0.09]$ & $[-0.05, 0.05]$ \\
\midrule

\multirow{4}{*}{$c_{\tilde{W}}/\Lambda^2$}
& \multirow{4}{*}{\onnWtil}
& no  & no  & $[-0.53, 0.54]$ & $[-0.74, 0.23]$ \\
& & yes & no  & $[-0.61, 0.36]$ & $[-0.67, 0.19]$ \\
& & no  & yes & $[-0.32, 0.32]$ & $[-0.45, 0.14]$ \\
& & yes & yes & $[-0.43, 0.25]$ & $[-0.54, 0.11]$ \\
\midrule

\multirow{4}{*}{$c_{HWB}/\Lambda^2$}
& \multirow{4}{*}{$\phi_f \times \pty$}
& no  & no  & $[-1.21, 0.90]$ & $[-0.67, 1.21]$ \\
& & yes & no  & $[-0.97, 0.82]$ & $[-0.47, 0.67]$ \\
& & no  & yes & $[-0.84, 0.62]$ & $[-1.09, 0.50]$ \\
& & yes & yes & $[-0.76, 0.60]$ & $[-0.58, 0.38]$ \\
\midrule

\multirow{4}{*}{$c_{H\tilde{W}B}/\Lambda^2$}
& \multirow{4}{*}{\onnHWBtil}
& no  & no  & $[-0.39, 0.39]$ & $[-0.59, 0.13]$ \\
& & yes & no  & $[-0.39, 0.39]$ & $[-0.60, 0.13]$ \\
& & no  & yes & $[-0.20, 0.21]$ & $[-0.31, 0.07]$ \\
& & yes & yes & $[-0.20, 0.21]$ & $[-0.31, 0.07]$ \\
\bottomrule
\end{tabular}
\end{table}

The expected and observed 95\% confidence intervals on the Wilson coefficients are reported in Table~\ref{tab:eft_limits}. The results are derived for each Wilson coefficient after setting the other Wilson coefficients to zero. Constraints are presented with and without the application of a $k$-factor to account for missing higher-order corrections in the EFT predictions. The constraints on the Wilson coefficients are up to a factor of two stronger with the $k$-factor applied, highlighting the importance of developing EFT predictions that include higher-order QCD effects for all SMEFT operators.  Constraints are also derived with and without the inclusion of the pure dimension-six term as a way to assess the impact of higher-order EFT contributions. Robust EFT interpretations are typically obtained when constraints on the Wilson coefficients are driven by the interference term, which is formally the leading deformation to the SM prediction as it is $O(\Lambda^{-2})$. The constraints on the CP-odd operators are largely unaffected by the inclusion of the pure-dimension six term, with \oHWBtil{} unchanged. Conversely, the constraints on the Wilson coefficient of the CP-even operator \oW{} tighten significantly  when the pure dimension-six contribution is included in the EFT prediction. The sensitivity to the Wilson coefficients is limited by statistical uncertainties.

The constraints on \cHWBtil{} represent a substantial improvement over previous measurements.
The most stringent prior limits on \cHWBtil{} were obtained using measurements of $H\rightarrow4\ell$ \cite{ATLAS:2023mqy} with 95\% confidence intervals of $[-0.72,\,0.72]$~TeV$^{-2}$ (expected) and $[-0.97,\,0.98]$~TeV$^{-2}$ (observed). Even without the application of a $k$-factor, the observed constraints on this operator are improved by a factor of 2.5 using the inclusive $W\gamma$ measurements presented in this paper. With the application of the $k$-factor, the observed constraints are improved by a factor of 5. The inclusion of the inclusive $W\gamma$ measurements in global fits for CP-violation in the Higgs sector~\cite{ATLAS:2026-071} would therefore improve the constraints on \cHWBtil{} considerably. For \cW{}, when using a $k$-factor to account for missing higher-order QCD corrections, the constraints are similar to both the constraints obtained by the CMS Collaboration in the $W\gamma$ final state~\cite{CMS:2021cxr} and to the constraints obtained in the ATLAS measurement of EW $Z\!j\!j$ production~\cite{ATLAS:2020nzk}.


\FloatBarrier
\section{Conclusions}
\label{sec:conc}

Differential cross sections for inclusive $W\gamma$ production have been measured in proton--proton collisions at $\sqrt{s}=13$~TeV using the full Run 2 ATLAS dataset of 140~fb$^{-1}$. The analysis is performed in the $W\gamma \rightarrow \ell\nu\gamma$ ($\ell = e,\mu$) channel and the measurements are unfolded to a particle-level fiducial phase space that closely matches the detector-level selection.

A comprehensive set of 16 observables is studied that collectively probe the kinematic
properties of the $W\gamma$ system, the radiation amplitude zero effect predicted in the $W\gamma$ final state, the polarisation of the $W$ boson, the charge conjugation and parity structure of the
$WW\gamma$ triple gauge coupling, and the parton distribution functions of the proton. The measurements achieve sufficient precision to discriminate between state-of-the-art theoretical predictions including higher-order NNLO QCD and (virtual) NLO EW corrections. Novel observables based on the output of neural networks are used to construct CP-sensitive observables that are highly sensitive to anomalous CP-violating $WW\gamma$ interactions. Furthermore, double differential cross sections are measured as a function of the $W\rightarrow \ell \nu$ decay angles, providing sensitivity to the $W$ boson spin density matrix. %

Within the framework of a dimension-six effective field theory, constraints are placed on the Wilson coefficients of the \oW, \oWtil, \oHWB{} and \oHWBtil{} operators using the differential cross section measurements. Constraints are placed on the Wilson coefficients of four operators: \oW{}, \oHWB{}, \oWtil{} and \oHWBtil{}.
The sensitivity to the \oHWBtil{} operator is improved by a factor of at least 2.5 relative to previous measurements in other final states.


%
%
%
%
%
\section*{Acknowledgements}
%

%

%
%

%
%

We thank CERN for the very successful operation of the LHC and its injectors, as well as the support staff at
CERN and at our institutions worldwide without whom ATLAS could not be operated efficiently.

The crucial computing support from all WLCG partners is acknowledged gratefully, in particular from CERN, the ATLAS Tier-1 facilities at TRIUMF/SFU (Canada), NDGF (Denmark, Norway, Sweden), CC-IN2P3 (France), KIT/GridKA (Germany), INFN-CNAF (Italy), NL-T1 (Netherlands), PIC (Spain), RAL (UK) and BNL (USA), the Tier-2 facilities worldwide and large non-WLCG resource providers. Major contributors of computing resources are listed in Ref.~\cite{ATL-SOFT-PUB-2026-001}.

We gratefully acknowledge the support of ANPCyT, Argentina; YerPhI, Armenia; ARC, Australia; BMWFW and FWF, Austria; ANAS, Azerbaijan; CNPq and FAPESP, Brazil; NSERC, NRC and CFI, Canada; CERN; ANID, Chile; CAS, MOST and NSFC, China; Minciencias, Colombia; MEYS CR, Czech Republic; DNRF and DNSRC, Denmark; IN2P3-CNRS and CEA-DRF/IRFU, France; SRNSFG, Georgia; BMFTR, HGF and MPG, Germany; GSRI, Greece; RGC and Hong Kong SAR, China; ICHEP and Academy of Sciences and Humanities, Israel; INFN, Italy; MEXT and JSPS, Japan; CNRST, Morocco; NWO, Netherlands; RCN, Norway; MNiSW, Poland; FCT, Portugal; MNE/IFA, Romania; MSTDI, Serbia; MSSR, Slovakia; ARIS and MVZI, Slovenia; DSI/NRF, South Africa; MICIU/AEI, Spain; SRC and Wallenberg Foundation, Sweden; SERI, SNSF and Cantons of Bern and Geneva, Switzerland; NSTC, Taipei; TENMAK, T\"urkiye; STFC/UKRI, United Kingdom; DOE and NSF, United States of America.

Individual groups and members have received support from BCKDF, CANARIE, CRC and DRAC, Canada; CERN-CZ, FORTE and PRIMUS, Czech Republic; COST, ERC, ERDF, Horizon 2020 and Marie Sk{\l}odowska-Curie Actions, European Union; Investissements d'Avenir Labex, Investissements d'Avenir Idex and ANR, France; DFG and AvH Foundation, Germany; Herakleitos, Thales and Aristeia programmes co-financed by EU-ESF and the Greek NSRF, Greece; BSF-NSF and MINERVA, Israel; NCN and NAWA, Poland; La Caixa Banking Foundation, CERCA and AGAUR programs from Generalitat de Catalunya and PROMETEO and GenT Programmes Generalitat Valenciana, Spain; G\"{o}ran Gustafssons Stiftelse, Sweden; The Royal Society and Leverhulme Trust, United Kingdom; Eric and Wendy Schmidt Fund for Strategic Innovation, United States of America.

In addition, individual members wish to acknowledge support from Chile: Agencia Nacional de Investigaci\'on y Desarrollo (ANID FONDECYT reg. 1230987, FONDECYT 1230812, FONDECYT 1240864, Fondecyt 3240661, Fondecyt Regular 1240721); China: Chinese Ministry of Science and Technology (MOST-2023YFA1605700, MOST-2023YFA1609300), National Natural Science Foundation of China (NSFC 12275265, NSFC-W2543005); Czech Republic: Czech Science Foundation (GACR - 24-11373S), Ministry of Education Youth and Sports (ERC-CZ-LL2327, FORTE CZ.02.01.01/00/22\_008/0004632), PRIMUS Research Programme (PRIMUS/21/SCI/017); EU: H2020 European Research Council (ERC - 101002463); European Union: European Research Council (BARD No. 101116429, ERC - 948254, ERC 101089007), European Regional Development Fund (HE COFUND GA No.101081355, ERDF), Marie Sklodowska-Curie Actions (GAP-101168829); France: Agence Nationale de la Recherche (ANR-21-CE31-0013, ANR-22-EDIR-0002, ANR-24-CE31-0504-01); Germany: Deutsche Forschungsgemeinschaft (DFG - 469666862); China: Research Grants Council (GRF); Italy: Ministero dell'Università e della Ricerca (NextGenEU 153D23001490006 M4C2.1.1, NextGenEU I53D23000820006 M4C2.1.1, NextGenEU I53D23001490006 M4C2.1.1, SOE2024\_0000023); Japan: Japan Society for the Promotion of Science (JSPS KAKENHI  JP25H0063, JSPS KAKENHI JP22H01227, JSPS KAKENHI JP22H04944, JSPS KAKENHI JP22KK0227, JSPS KAKENHI JP24K23939, JSPS KAKENHI JP24KK0251, JSPS KAKENHI JP25H00650, JSPS KAKENHI JP25H01291, JSPS KAKENHI JP25K01023); Poland: Polish National Science Centre (NCN 2021/42/E/ST2/00350, NCN OPUS 2023/51/B/ST2/02507, NCN OPUS nr 2022/47/B/ST2/03059, NCN UMO-2019/34/E/ST2/00393, UMO-2022/47/O/ST2/00148, UMO-2023/49/B/ST2/04085, UMO-2023/51/B/ST2/00920, UMO-2024/53/N/ST2/00869); Spain: Agència de Gestió d'Ajuts Universitaris i de Recerca. (AGAUR - 2023 BP 00141), Ministry of Science and Innovation (RYC2019-028510-I, RYC2020-030254-I, RYC2021-031273-I, RYC2022-038164-I), Ministerio de Ciencia, Innovación y Universidades/Agencia Estatal de Investigaci\'on (PID2022-142604OB-C22); Sweden: Carl Trygger Foundation (Carl Trygger Foundation CTS 22:2312), Swedish Research Council (Swedish Research Council 2023-04654, VR 2021-03651, VR 2022-03845, VR 2022-04683, VR 2023-03403, VR 2024-05451, VR 2025-05940), Knut and Alice Wallenberg Foundation (KAW 2023.0366); Switzerland: Swiss National Science Foundation (SNSF - PCEFP2\_194658); United Kingdom: The Binks Trust, Royal Society (NIF-R1-231091); United States of America: U.S. Department of Energy (ECA DE-AC02-76SF00515), John Templeton Foundation (John Templeton Foundation 63206), Neubauer Family Foundation.

%
%


%
\clearpage
\appendix
\part*{Appendix}
\addcontentsline{toc}{part}{Appendix}

Figure \ref{fig:diffxs1a} shows the differential cross sections measured as a function of \ptl{}, \mtlvy{} and \dphily{}. The \ptl{} and \mtlvy{} distributions are measured in the \textrm{baseline} region, whereas the \dphily{} measurement is performed in the \textit{jet veto} region. Figure \ref{fig:polarisation_observables2} presents the  double differential cross section as a function of $\phi_f$ and \pty{} in the \textit{baseline} fiducial region. Figure~\ref{fig:boost_asymmetry2} shows the boost asymmetry measured as a function of $|\eta_\ell|$. In all cases, the agreement between data and the theoretical predictions is similar to the observables shown in the main body of the paper.

\begin{figure}[hbp]
\centering
\subfloat[]{\includegraphics[width=0.45\textwidth]{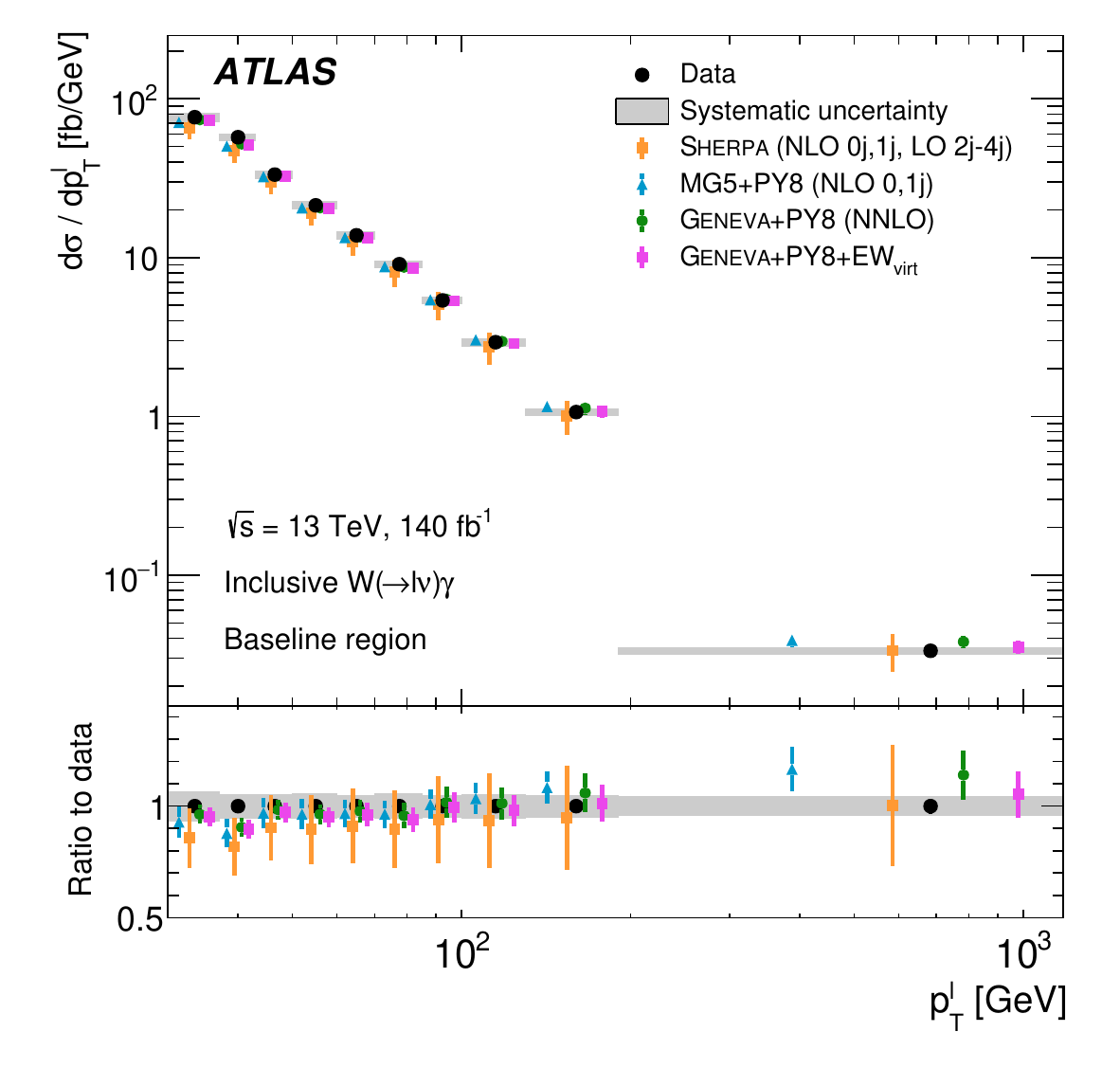}}
\subfloat[]{\includegraphics[width=0.45\textwidth]{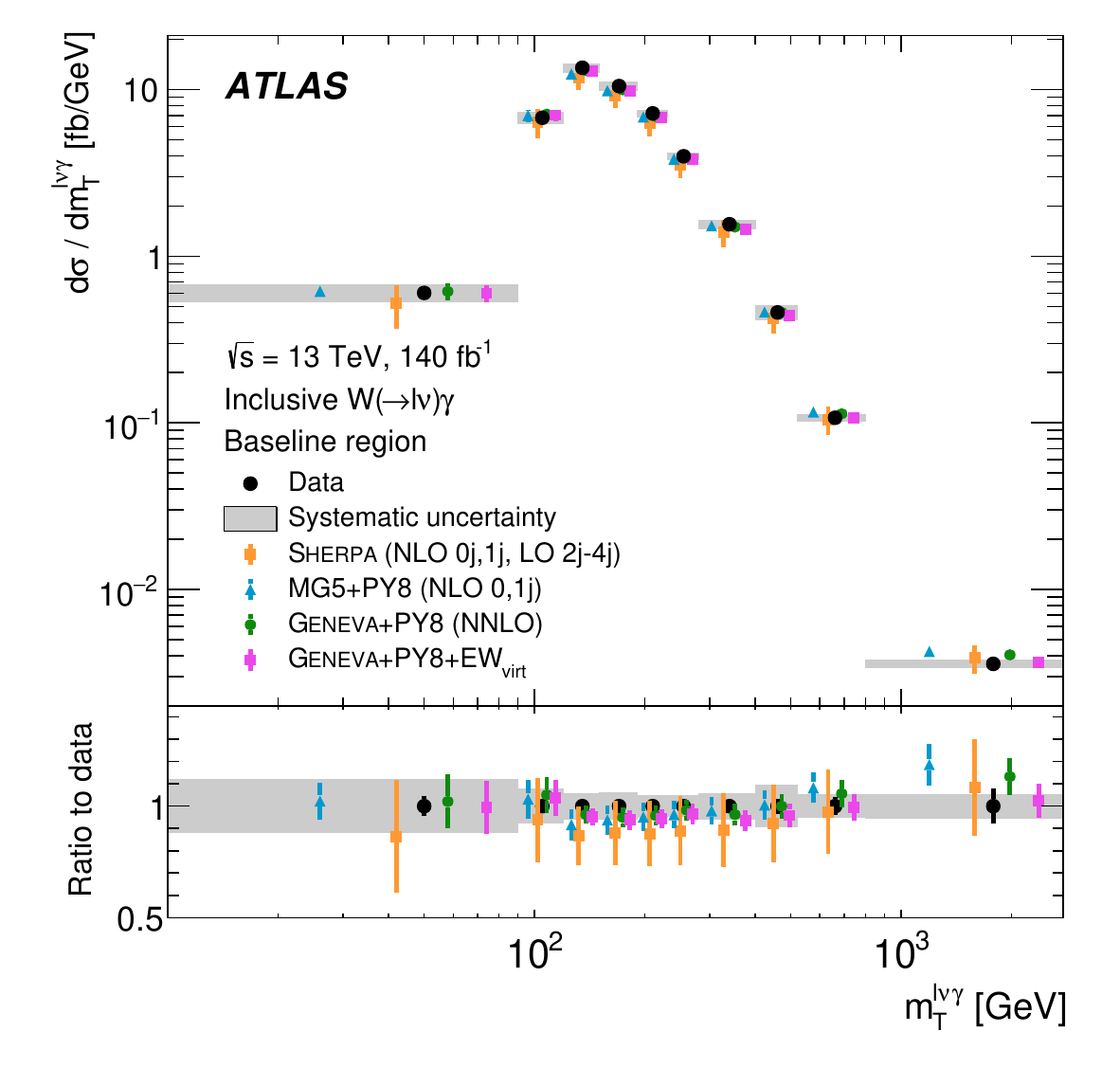}}\\
\subfloat[]{\includegraphics[width=0.45\textwidth]{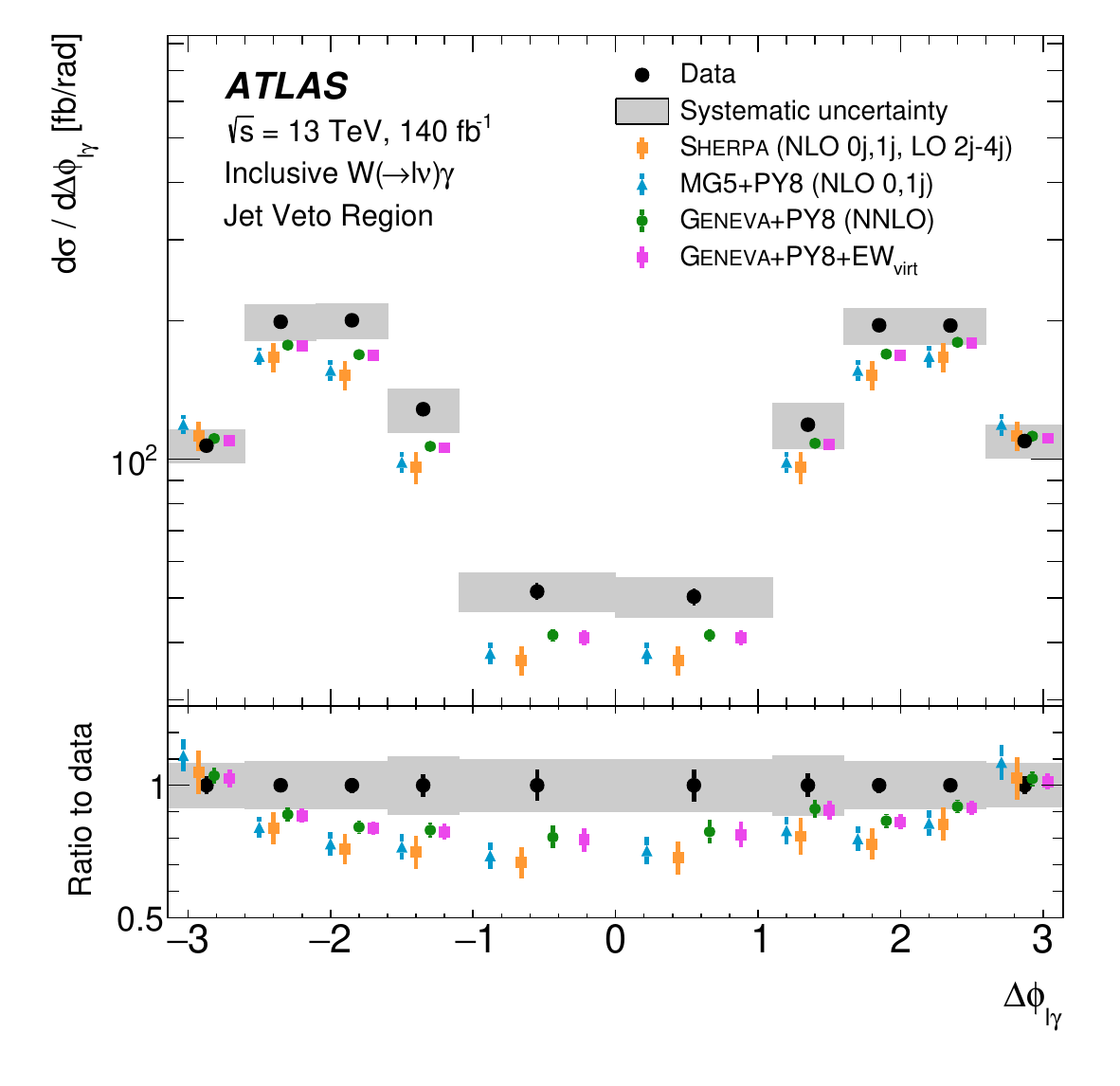}}
\caption{Unfolded differential cross sections measured as a function of
(a)  \ptl{}, (b) \mtlvy{} and (c) \dphily{}.The data and theoretical predictions are presented in the same way as for Figure~\ref{fig:diff_xs1}.}
\label{fig:diffxs1a}
\end{figure}

\begin{figure}[hbp]
\centering
\includegraphics[width=0.99\textwidth]{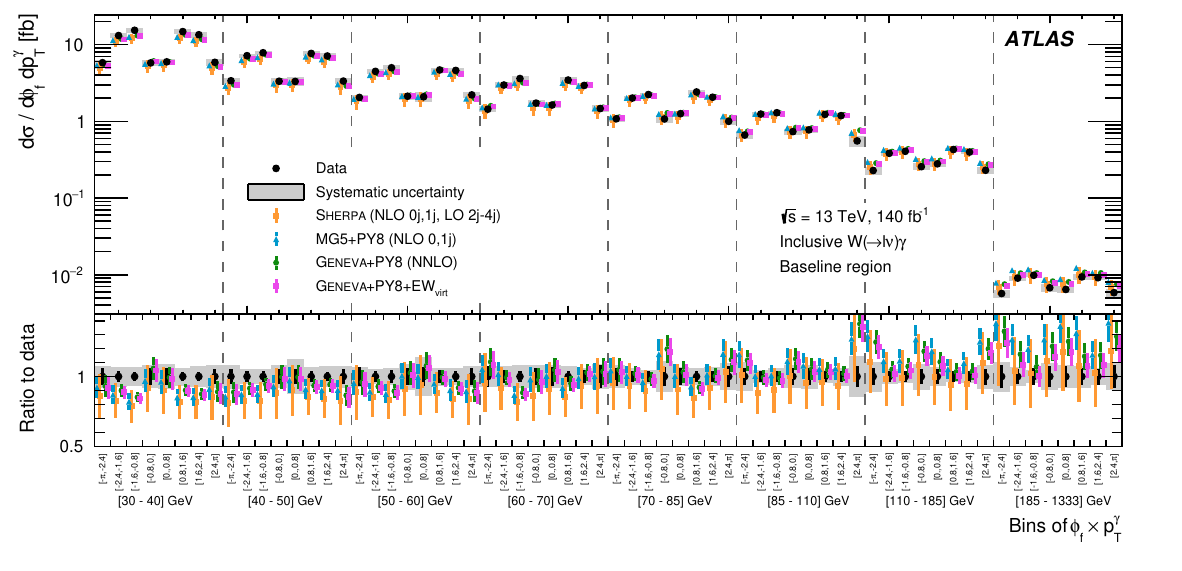}
\caption{Double differential cross section of inclusive $W\gamma$ production as a function of $\phi_f$ and \pty{} in the \textrm{baseline} fiducial region. The data and theoretical predictions are presented in the same way as for Figure~\ref{fig:polarisation_observables1}.}
\label{fig:polarisation_observables2}
\end{figure}

\begin{figure}[hbp]
\centering
\subfloat[]{\includegraphics[width=0.45\textwidth]{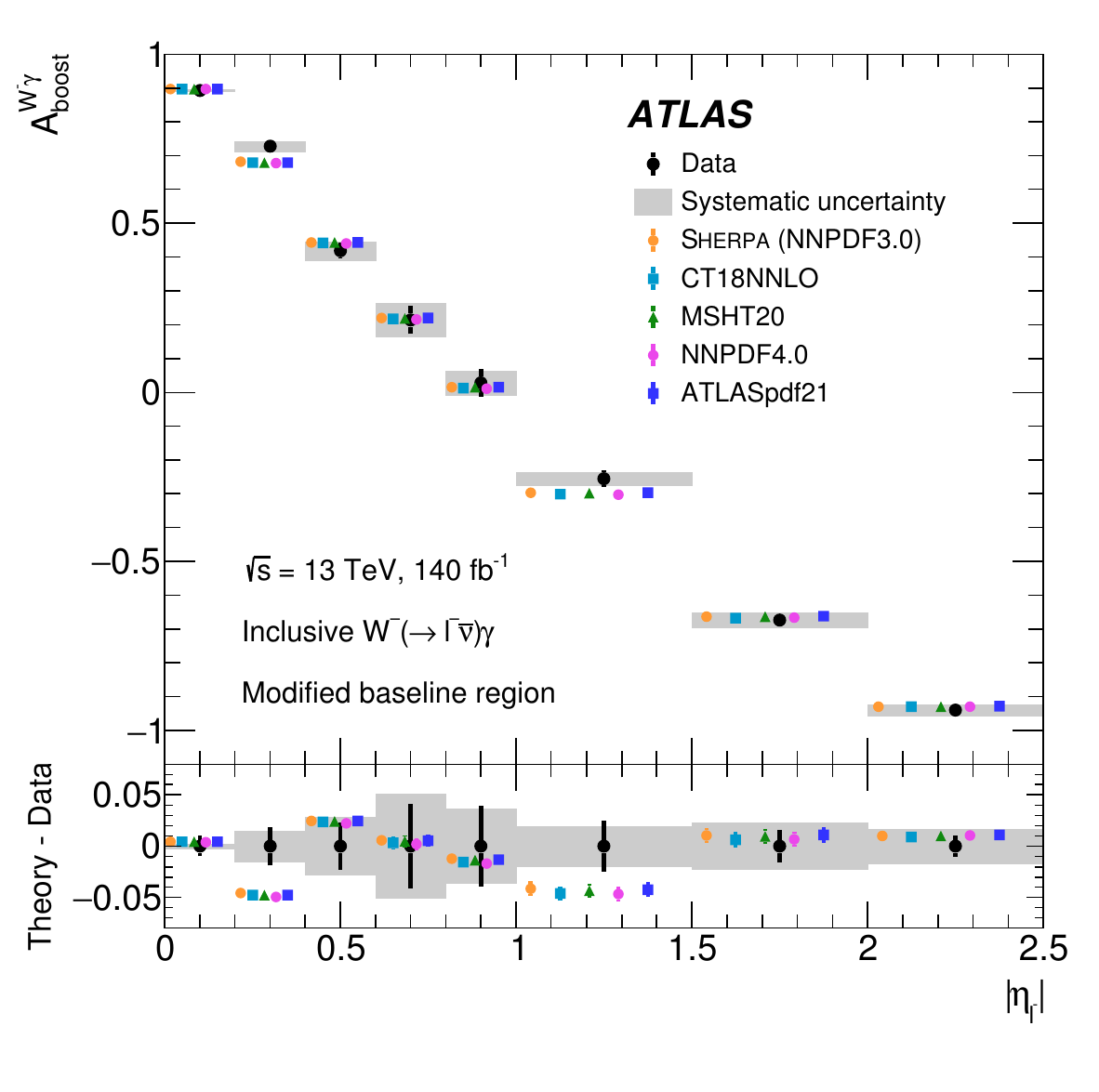}}
\subfloat[]{\includegraphics[width=0.45\textwidth]{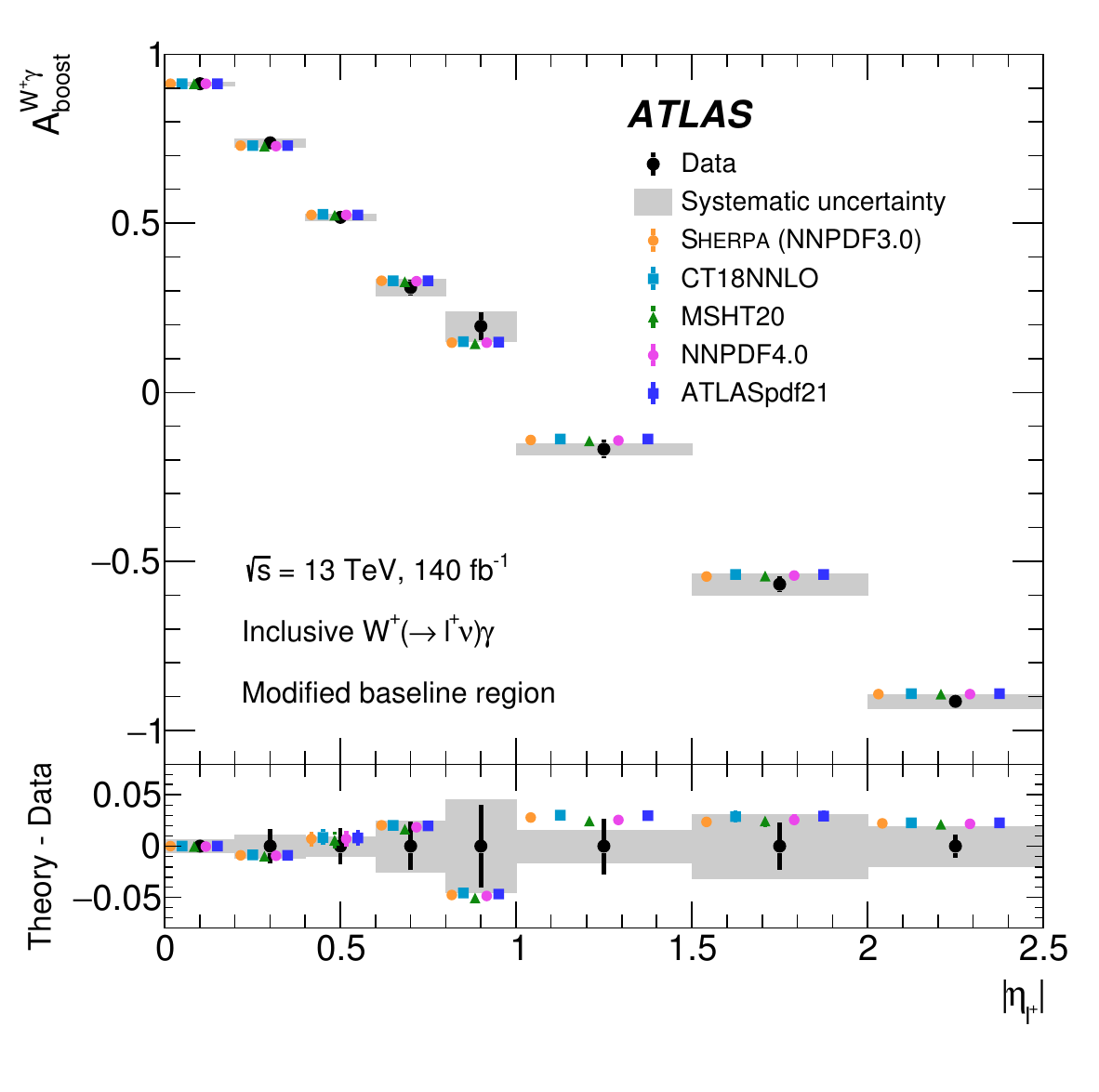}}
\caption{Boost asymmetry measured as a function of $|\eta_{\ell}|$ for (a) $W^{-}\gamma$ production and (b) $W^{+}\gamma$ production (b). The data and theoretical predictions are presented in the same way as for Figure~\ref{fig:boost_asymmetry}.}
\label{fig:boost_asymmetry2}
\end{figure}


\label{sec:app}
\printbibliography
\clearpage
 
\begin{flushleft}
\hypersetup{urlcolor=black}
{\Large The ATLAS Collaboration}

\bigskip

\AtlasOrcid[0000-0002-6665-4934]{G.~Aad}$^\textrm{\scriptsize 102}$,
\AtlasOrcid[0000-0001-7616-1554]{E.~Aakvaag}$^\textrm{\scriptsize 17}$,
\AtlasOrcid[0000-0002-5888-2734]{B.~Abbott}$^\textrm{\scriptsize 121}$,
\AtlasOrcid[0000-0002-0287-5869]{S.~Abdelhameed}$^\textrm{\scriptsize 83b}$,
\AtlasOrcid[0000-0002-1002-1652]{K.~Abeling}$^\textrm{\scriptsize 54}$,
\AtlasOrcid[0000-0001-5763-2760]{N.J.~Abicht}$^\textrm{\scriptsize 48}$,
\AtlasOrcid[0000-0002-8496-9294]{S.H.~Abidi}$^\textrm{\scriptsize 30}$,
\AtlasOrcid[0009-0003-6578-220X]{M.~Aboelela}$^\textrm{\scriptsize 44}$,
\AtlasOrcid[0000-0002-9987-2292]{A.~Aboulhorma}$^\textrm{\scriptsize 36e}$,
\AtlasOrcid[0000-0001-5329-6640]{H.~Abramowicz}$^\textrm{\scriptsize 154}$
% ...
\bigskip
\\

% ...
$^{17}$Department for Physics and Technology, University of Bergen, Bergen; Norway.\\
% ...
$^{30}$Physics Department, Brookhaven National Laboratory, Upton NY; United States of America.\\
% ...
$^{36}$$^{(a)}$Facult\'e des Sciences Ain Chock, Universit\'e Hassan II de Casablanca;$^{(b)}$Facult\'{e} des Sciences, Universit\'{e} Ibn-Tofail, K\'{e}nitra;$^{(c)}$Facult\'e des Sciences Semlalia, Universit\'e Cadi Ayyad, LPHEA-Marrakech;$^{(d)}$LPMR, Facult\'e des Sciences, Universit\'e Mohamed Premier, Oujda;$^{(e)}$Facult\'e des sciences, Universit\'e Mohammed V, Rabat;$^{(f)}$Institute of Applied Physics, Mohammed VI Polytechnic University, Ben Guerir; Morocco.\\
% ...
$^{44}$Physics Department, Southern Methodist University, Dallas TX; United States of America.\\
% ...
$^{48}$Fakult\"{a}t Physik, Technische Universit{\"a}t Dortmund, Dortmund; Germany.\\
% ...
$^{54}$II. Physikalisches Institut, Georg-August-Universit\"{a}t G\"ottingen, G\"ottingen; Germany.\\
% ...
$^{83}$$^{(a)}$Khalifa University of Science and Technology, Abu Dhabi;$^{(b)}$New York University Abu Dhabi, Abu Dhabi;$^{(c)}$United Arab Emirates University, Al Ain;$^{(d)}$University of Sharjah, Sharjah; United Arab Emirates.\\
% ...
$^{102}$CPPM, Aix-Marseille Universit\'e, CNRS/IN2P3, Marseille; France.\\
% ...
$^{121}$Homer L. Dodge Department of Physics and Astronomy, University of Oklahoma, Norman OK; United States of America.\\
% ...
$^{154}$Raymond and Beverly Sackler School of Physics and Astronomy, Tel Aviv University, Tel Aviv; Israel.\\
% ...
$^{a}$ Also at Affiliated with an institute formerly covered by a cooperation agreement with CERN.\\
$^{b}$ Also at An-Najah National University, Nablus; Palestine.\\
$^{c}$ Also at Borough of Manhattan Community College, City University of New York, New York NY; United States of America.\\
$^{d}$ Also at Center for Interdisciplinary Research and Innovation (CIRI-AUTH), Thessaloniki; Greece.\\
$^{e}$ Also at Centre of Physics of the Universities of Minho and Porto (CF-UM-UP); Portugal.\\
$^{f}$ Also at CERN, Geneva; Switzerland.\\
% ...

\end{flushleft}


%

%
%

%

%
%
%
%

\end{document}